\documentstyle[aps,prb]{revtex}

\begin{document}
\draft
\title{Electro-optical properties of semiconductor quantum dots: 
Application to quantum information processing}
\author{Eliana Biolatti$^{1,2}$, Irene D'Amico$^{1,3}$, 
Paolo Zanardi$^{1,3}$, and Fausto Rossi$^{1,2,3}$}
\address{$^1$ Istituto Nazionale per la Fisica della Materia (INFM)}
\address{$^2$ Dipartimento di Fisica, Politecnico di Torino,
Corso Duca degli Abruzzi 24 \\
I-10129 Torino, Italy}
\address{$^3$ Institute for Scientific Interchange (ISI), 
Villa Gualino, Viale Settimio Severo 65, I-10133 Torino, Italy}
\maketitle

\begin{abstract}

A detailed analysis of the electro-optical response of single as 
well as coupled semiconductor quantum dots is presented. This is based on a
realistic ---i.e., fully tridimensional--- description of 
Coulomb-correlated few-electron 
states, obtained via a direct-diagonalization approach.
More specifically,
we investigate the combined effect of static electric fields and ultrafast 
sequences of multicolor laser pulses in the few-carrier, i.e., low-excitation, 
regime. 
In particular, we show how the presence of a properly tailored static 
field may give rise to significant electron-hole charge separation; 
these field-induced dipoles, in turn, may introduce relevant 
exciton-exciton couplings, which are found to induce significant 
---both intra- and inter-dot--- biexcitonic splittings.
We finally show that such few-exciton systems constitute an ideal 
semiconductor-based hardware for an {\it all optical} implementation of 
quantum information processing.

\end{abstract}


\clearpage

\section{Introduction}\label{s:int}

In the past years increasing interest has been focused on  semiconductor
nanostructures.\cite{SN} This is mainly due to their low-dimensional 
character, which allows to tailor carrier quantum confinement as well as 
Coulomb interaction. As a result, this has allowed to fabricate nanostructured 
systems with a properly designed density-of-states which, in turn, 
exhibit an increased optical efficiency as well as a reduction of  
energy-relaxation and dephasing processes.\cite{SST} 
For the case of two- and one-dimensional nanostructures, i.e., quantum 
wells and wires, however, we deal with a partial carrier confinement, i.e.,
the single-particle energy spectrum is still continuous. 
This allows to describe their many-body ultrafast optical response in terms
of the usual mean-field approaches, typical of bulk systems.\cite{Kuhn98}

The real scientific and technological ``revolution'' in the field was the 
introduction of quasi zero-dimensional (0D) systems, called semiconductor 
quantum dots.\cite{QDs}
Compared to systems of higher dimensionality ---like quantum wells and 
wires--- they have a discrete, i.e., atomic-like, energy spectrum and, more 
important, they exhibit genuine few-carrier effects. 
Generally speaking, going from quantum wells and wires to quantum dots 
(QDs) we move from many-electron systems to few-electron ones.
This implies a radical change in the theoretical schemes\cite{QD-theo} 
as well as in the experimental techniques\cite{QD-exp} used to study such 
quasi 0D nanostructures, often 
referred to as {\it semiconductor macroatoms}.
Apart from their relevance in terms of basic physics, 
these novel semiconductor nanostructures have attracted 
general attention because of  their technological
applications: these range from laser emitters\cite{laser} 
 to 
charge-storage devices,\cite{stor} from fluorescent biological 
markers\cite{bio} 
to quantum information processing devices.\cite{QIPD} 

In QD's,  the flexibility
typical of  semiconductors in controlling carrier densities has been
brought to its extreme:   it is possible to electrically inject  single
electrons\cite{Coulblock} or to photogenerate in a QD  
a single Coulomb-correlated electron-hole pair, i.e., a single 
exciton.\cite{excit,Hawry} 
It is even  possible to detect the 
single-exciton decaying energy emission.\cite{excit,Hawry} 
The quantized,
atomic-like, energy structure of QD's allows for a rich optical spectrum
and for a weak interaction of the QD system with environmental degrees of
freedom (like phonons, plasmons, etc.). This latter feature implies   
that the  quantum evolution of the
carrier subsystem is affected by low decoherence.\cite{ZR}

Moreover, their reduced spatial extension ---up to few nanometers--- 
leads to an increase of two-body interactions among carriers and to stronger 
Coulomb-correlation effects.\cite{QD-theo}
The latter may be used to design a variety of {\it single-electron 
devices}. In particular, as we shall show, they can be employed to design 
{\it fully-optical quantum gates}, as recently proposed in 
Ref.~\onlinecite{PRL}.
Indeed, the continuous progress in QD fabrication and 
characterization\cite{manip} let us  foresee a near future in which 
it will be possible to exactly tailor the few-carrier and optical 
properties  of these 0D systems. 
In this respect, a step forward has been recently made
by the analysis  and understanding of a single-QD excitonic
emission spectrum,\cite{Hawry,HawryPRB} that uncovered ``hidden''
symmetries in isolated QD structures, analogous to  Hund's rules\cite{Hawry}
for real atoms. These symmetries imply that, under suitable conditions,
Coulomb correlations among excitons in the same dot cancel. 

The primary goal of this paper is twofold. 
On the one hand, we shall present a detailed investigation of the 
electro-optical response of single- as well as coupled QD structures. More 
specifically, we shall focus on the combined effect of static electric 
fields and ultrafast multicolor laser pulses. Our investigation will 
present a variety of field-induced effects unexplored so far; in particular,
we shall show how a properly tailored external field can be used to induce or 
reinforce exciton-exciton Coulomb coupling both in single and coupled QD 
structures.
On the other hand, we shall discuss the application of such field-induced 
few-exciton effects to design a {\it semiconductor-based fully-optical 
quantum information processing strategy}.\cite{PRL}

The paper is organized as follows: 
In Sect.~\ref{s:ta} we shall introduce our theoretical
approach for the analysis of the electro-optical response of QD structures; 
Section \ref{s:eor} presents a detailed investigation of the excitonic as 
well as biexcitonic response of 
prototypical semiconductor macroatoms and 
molecules in the presence of an applied static field;
In Sect.~\ref{s:qip} our quantum information processing 
strategy is discussed and a few simulated experiments of basic 
quantum information/computation (QIC) operations are presented;
Finally, in Sect.~\ref{s:conc} we shall summarize and draw some 
conclusions.

\section{Theoretical approach}\label{s:ta}

The physical system under investigation is a gas of electron-hole pairs 
confined in a
quasi-0D semiconductor structure, i.e., a single as well as a multiple 
QD. 
In this case, the total Hamiltonian of our semiconductor nanostructure 
can be regarded as the sum of two terms, 
${\bf H} = {\bf H}^\circ + {\bf H}'$: 
A term ${\bf H}^\circ$ describing the
correlated electron-hole subsystem, i.e., 
free carriers plus confinement potential plus carrier-carrier Coulomb 
interaction,
and a term ${\bf H}'$ describing the interaction of the carrier subsystem with 
coherent-light sources and environmental degrees of freedom, i.e., 
carrier-light plus carrier-phonon interactions.

\subsection{Single-particle description}\label{ss:spd}

Let us first consider the gas of non-interacting carriers, electrons ($e$) and
holes ($h$) confined within the quasi-0D semiconductor structure. The
quantum confinement can be described in terms of an effective potential 
$V_c^{e/h}$ whose
height is dictated by the conduction/valence band discontinuities. 
Since the energy region of interest is relatively close to the band gap 
$\epsilon_{{\rm gap}}$ of the semiconductors forming our heterostructure,
we shall describe the bulk band structure in terms of the usual effective-mass
approximation.\cite{Cardona} In addition, since the confinement potential
$V_c^{e/h}$ is a slowly-varying function on the scale of the lattice
periodicity, we shall work within the ``envelope-function'' 
picture.\cite{Bastard}

Within such approximation scheme, the non-interacting carriers in our quasi-0D
structure are then described by the following Schr\"odinger equation 
\begin{equation}\label{Schr1}
\left[
-{\hbar^2\nabla^2_{\bf r}\over 2 m_{e/h}} 
+ V_c^{e/h}({\bf r})
\right] \psi_{i/j}({\bf r}) = \epsilon_{i/j} \psi_{i/j}({\bf r}) \ ,
\end{equation}
where $m_{e/h}$ is the bulk effective mass for electrons/holes 
while $i/j$ denotes the set of single-particle quantum numbers, including 
charge as well as spin degrees of freedom.\cite{spin}
Here, $\psi_{i/j}({\bf r})$ is the envelope function of state $i (j)$,
 the eigenvalues $\epsilon_{i/j}$ correspond to the energy
levels of the carriers induced by the confinement-potential profile
$V_c^{e/h}$; since the latter ---for any realistic semiconductor 
nanostructure---
is finite, the lowest part of the single-particle energy spectrum 
$\epsilon_{i/j}$ is discrete, while for increasing energies it evolves into
a continuum.
The different approaches commonly employed for the solution of 
Eq.~(\ref{Schr1}) are described in App.~\ref{app:spp}; 
according to the energy region of interest, they range from direct 
three-dimensional (3D) plane-wave expansions, to factorized-state solutions, or to 
simplified two-dimensional (2D) parabolic-potential models.

\subsection{Coulomb-correlated carrier system}\label{ss:cccs}

Given the above single-particle representation $\{ i \}$ ($\{ j \}$) for 
electrons (holes)
---i.e., the set of 3D eigenfunctions 
$\psi_i({\bf r}) \equiv \langle {\bf r} \vert i \rangle$
($\psi_j({\bf r}) \equiv \langle {\bf r} \vert j \rangle$)
and the corresponding energy levels
$\epsilon_i$ ($\epsilon_j$)---
let us now introduce the following creation and destruction operators for 
electrons and holes:
\begin{equation}\label{cre-des}
\vert i \rangle = c^\dagger_i \vert 0 \rangle
\to
\vert 0 \rangle = c^{ }_i \vert i \rangle \ , \quad
\vert j \rangle = d^\dagger_j \vert 0 \rangle
\to
\vert 0 \rangle = d^{ }_j \vert j \rangle \ ,
\end{equation}
where $\vert 0 \rangle$ denotes the electron-hole vacuum state.
Within such second-quantization picture, 
the single-particle Hamiltonian, i.e., the Hamiltonian describing the 
non-interacting carriers within our 0D confinement potential,
can be written as: 
\begin{equation}\label{H_c}
{\bf H}^c = {\bf H}^e + {\bf H}^h =
\sum_i \epsilon_i c^\dagger_i c^{ }_i + 
\sum_j \epsilon_j d^\dagger_j d^{ }_j \ .
\end{equation}

The carriers (electrons and holes) within our quasi-0D nanostructure, 
however, interact via the two-body Coulomb potential $V({\bf r-r'})$. 
Due to such interaction, several correlation effects
take place. Here, only processes conserving the total number of carriers
are considered, thus Auger recombination and impact ionization are
neglected. Such processes are known to become important only at very high
densities and at energies high up in the band.\cite{II} 
In this case the Hamiltonian describing carrier-carrier interaction within our 
single-particle $i/j$-picture can be written as
\begin{eqnarray}\label{H_cc}
{\bf H}^{cc} &=& {\bf H}^{ee} + {\bf H}^{hh} + {\bf H}^{eh} \nonumber \\
& = & \frac{1}{2} 
\sum_{i_1, i_2, i_3, i_4}
V_{i_1 i_2 i_3 i_4} \,
c^\dagger_{i_1} c^\dagger_{i_2} c^{ }_{i_3} c^{ }_{i_4}
\nonumber \\
& + & \frac{1}{2} 
\sum_{j_1, j_2, j_3, j_4}
V_{j_1 j_2 j_3 j_4} \,
d^\dagger_{j_1} d^\dagger_{j_2} d^{ }_{j_3} d^{ }_{j_4}
\nonumber \\
& - & 
\phantom{\frac{1}{2}}
\sum_{i_1, i_2, j_1, j_2}
V_{i_1 j_1 j_2 i_2} \,
c^\dagger_{i_1}
d^\dagger_{j_1}
d^{ }_{j_2}
c^{ }_{i_2}\ ,
\end{eqnarray}
where 
\begin{equation}\label{V_cc}
V_{l_1' l_2' l_2 l_1} = 
\int d{\bf r} \int d{\bf r'} 
\psi^*_{l_1'}({\bf r})
\psi^*_{l_2'}({\bf r'})
V({\bf r-r'})
\psi^{ }_{l_2}({\bf r'})
\psi^{ }_{l_1}({\bf r})
\end{equation}
are the matrix elements of the Coulomb potential 
for the generic two-particle transition 
$l_1 l_2 \to l_1' l_2'$.
The first two terms on the rhs of Eq.~(\ref{H_cc}) describe the repulsive
electron-electron and hole-hole interactions while the third one describes
the attractive interaction between electrons and holes. 

We stress the full 3D nature of the present approach based 
on the detailed knowledge of the 3D carrier wavefunctions 
$\psi$.
The explicit evaluation of the above matrix elements 
for a generic 
3D confinement-potential profile $V_c^{e/h}$, i.e., for a generic set 
of envelope functions $\psi_{i/j}$, is described in App.~\ref{app:cme}. 

Combining the single-particle Hamiltonian in (\ref{H_c}) with the 
Coulomb-interaction term in (\ref{H_cc}), we get the following many-body 
Schr\"odinger equation for our Coulomb-correlated system:
\begin{equation}\label{Schr2}
{\bf H}^\circ \vert \Psi \rangle = \left({\bf H}^c + {\bf H}^{cc}\right) 
\vert \Psi \rangle = {\cal E} \vert \Psi \rangle \ .
\end{equation}
Here, $\vert \Psi \rangle$ denotes the interacting many-body state 
in our Fock space and ${\cal E}$ the corresponding total energy.

Let us now introduce the total-number operators for electrons and holes:
\begin{equation}\label{N_eh}
{\bf N}_e = \sum_i c^\dagger_i c^{ }_i \ , \quad
{\bf N}_h = \sum_j d^\dagger_j d^{ }_j \ .
\end{equation}
It is easy to show that the above global operators commute with the 
carrier Hamiltonian ${\bf H}^\circ$ in (\ref{Schr2}).
We can therefore look for many-body states $\vert \Psi \rangle$ 
corresponding to a given number of electrons ($N_e$) and holes ($N_h$).
In particular, we shall consider the case of intrinsic semiconductor 
materials,\cite{Cardona} 
for which $N_e = N_h$; in this case the good quantum number is 
the total number of electron-hole pairs: $N = N_e = N_h$ and the 
Schr\"odinger equation (\ref{Schr2}) can be rewritten as:
\begin{equation}\label{Schr3}
{\bf H}^\circ \vert \lambda_N \rangle = 
{\cal E}_{\lambda_N} \vert \lambda_N \rangle \ ,
\end{equation}
where $\vert \lambda_N \rangle$ and 
${\cal E}_{\lambda_N}$ 
denote, respectively, the $\lambda$-th 
many-body state and energy level
corresponding to $N$ electron-hole pairs. 

For any given number $N$ of electron-hole pairs we thus identify a subspace
of the original Fock space, for which there exists a natural basis 
$\left\{\vert l_N \rangle\right\}$, given by the 
eigenstates of the single-particle Hamiltonian in (\ref{H_c}):
\begin{equation}\label{Schr4}
{\bf H}^c \vert l_N \rangle = 
\epsilon_{l_N} \vert l_N \rangle \ .
\end{equation}
Here, 
$l_N \equiv i_1,i_2,\dots,i_N; j_1,j_2,\dots,j_N$ is a compact notation for 
the set of non-interacting electron and hole single-particle quantum numbers 
corresponding to our $N$ electron-hole pairs.
Indeed, we have:
\begin{equation}\label{ket_lN}
\vert l_N \rangle \equiv 
\vert \{i_n j_n\} \rangle =
\prod_{n = 1}^N c^\dagger_{i_n} d^\dagger_{j_n} \vert 0 \rangle
\end{equation}
and
$
\epsilon_{l_N} = \sum_{n = 1}^N 
\left(\epsilon_{i_n} + \epsilon_{j_n} \right) 
$.

The non-interacting basis set in (\ref{ket_lN}) constitutes the starting point 
of the direct-diagonalization approach used for the solution of the many-body 
Schr\"odinger equation (\ref{Schr3}).
Indeed, we can expand the unknown many-body state 
$\vert \lambda_N \rangle$ over our single-particle basis:
\begin{equation}\label{ket_lambdaN}
\vert \lambda_N \rangle = \sum_{l_N} U^{\lambda_N}_{ l_N} 
\vert l_N \rangle \ .
\end{equation}
By inserting the above expansion into Eq.~(\ref{Schr3}), 
the latter is transformed into the following eigenvalue problem:
\begin{equation}\label{ep}
\sum_{l'_N} \left(H^\circ_{l^{ }_N l'_N} - {\cal E}_{\lambda_N} 
\delta_{l^{ }_N l'_N}\right) 
U^{\lambda^{ }_N}_{l'_N} = 0 
\ ,
\end{equation}
where
\begin{equation}\label{H_llprime}
H^\circ_{l^{ }_N l'_N}
= \langle l^{ }_N \vert  
{\bf H}^\circ
\vert l'_N \rangle =
\epsilon_{l_N} \delta_{l^{ }_N l'_N} + {\cal V}_{l^{ }_N l'_N}
\end{equation}
are the matrix elements of the carrier Hamiltonian ${\bf H}^\circ$ in our 
single-particle basis. They are given by a diagonal ---i.e., 
non-interacting--- contribution plus a non-diagonal term given by the matrix
elements of the Coulomb-interaction Hamiltonian in (\ref{H_cc}):
$
{\cal V}_{l^{ }_N l'_N} = \langle l^{ }_N \vert 
{\bf H}_{cc}
\vert l'_N \rangle 
$.
Their explicit form ---which involves the various two-body Coulomb matrix 
elements in (\ref{V_cc})--- is given in App.~\ref{app:emb} for the 
excitonic ($N = 1$) and biexcitonic ($N = 2$) case.

In the presence of Coulomb interaction, the Hamiltonian matrix in
(\ref{H_llprime}) is non-diagonal; therefore, the interacting many-body 
states $\vert \lambda_N \rangle$ are, in general, a linear 
superposition of all the single-particle states 
$\vert l_N \rangle$ [see Eq.~(\ref{ket_lambdaN})], 
whose coefficients $U^{\lambda_N}_{l_N}$ can 
be regarded as elements of the unitary transformation connecting the 
single-particle to the interacting basis:
$U^{\lambda_N}_{l_N} = \langle l_N \vert \lambda_N \rangle$.

The numerical evaluation of our Coulomb-correlated states is thus performed
by direct diagonalization of the Hamiltonian matrix $H^\circ$ in 
(\ref{H_llprime}), using a large ---but finite--- single-particle basis set.

\subsection{Interaction with coherent light sources}\label{ss:cl}

The Coulomb-correlated carrier system described so far will interact 
strongly with electromagnetic fields in the optical range.
For the case of a coherent light source ---the one considered in this 
paper--- the light-matter interaction Hamiltonian in our 
second-quantization picture can be written as:
\begin{equation}\label{H_prime}
{\bf H}' =
- E(t)
\sum_{ij}
\left[
\mu^{*}_{ij}
c^\dagger_i d^\dagger_j
+ 
\mu^{ }_{ij}
d^{ }_j c^{ }_i
\right] \ ,
\end{equation}
where $E(t)$ is the classical light-field, and
\begin{equation}\label{mu}
\mu_{ij} = \mu_{bulk} \int \psi_i({\bf r}) \psi_j({\bf r}) d {\bf r}
\end{equation}
is the dipole matrix element for the $ij$ transition, $\mu_{bulk}$ being 
its bulk value.
In the presence of a time-dependent coherent optical excitation the 
quantum-mechanical evolution of our electron-hole system will be described 
by the following time-dependent Schr\"odinger equation:
\begin{equation}\label{td-Schr1}
i \hbar \frac{d}{dt} \vert \Psi(t) \rangle = {\bf H} \vert \Psi(t) \rangle = 
\left({\bf H}^\circ + {\bf H}'\right) \vert \Psi(t) \rangle\ .
\end{equation}
Contrary to the carrier Hamiltonian ${\bf H}^\circ$, the carrier-light term
${\bf H}'$ does not commute with the global number operators in 
(\ref{N_eh}). Indeed, the two terms in (\ref{H_prime}) describe, respectively,
the light-induced creation and destruction of an electron-hole pair.
Therefore, $N$ is no more a good quantum number and the many-body state 
at time $t$ is, in general, a linear superposition of all 
the correlated $N$-pair basis states:
\begin{equation}\label{Psi}
\vert\Psi(t)\rangle = \sum_N \sum_{\lambda_N} a_{\lambda_N}(t) 
\vert \lambda_N \rangle \ .
\end{equation}
By inserting the above expansion into the time-dependent Schr\"odinger 
equation (\ref{td-Schr1}) we get: 
\begin{equation}\label{td-Schr2}
i \hbar \frac{d}{dt} a_{\lambda_N}(t) = { \cal E}_{\lambda_N} a_{\lambda_N}(t)+
\sum_{N'} \sum_{\lambda'_{N'}} 
H'_{\lambda^{ }_N \lambda'_{N'}} a_{\lambda'_{N'}}(t) \ ,
\end{equation}
where
\begin{equation}\label{H_lambda_lambdaprime}
H'_{\lambda^{ }_N \lambda'_{N'}} = 
\langle \lambda^{ }_N \vert {\bf H}' \vert \lambda'_{N'} \rangle
\end{equation}
are the matrix elements of the light-matter Hamiltonian (\ref{H_prime}) 
within our interacting $N$-pair basis $\{\lambda_N\}$.

It can be easily shown (see App.~\ref{app:emb}) 
that the above matrix elements are different from zero only
for $N' = N \pm 1$; this confirms that the only possible transitions are $N
\to N+1$ or $N+1 \to N$ which correspond, respectively,  to the generation 
and destruction
of Coulomb-correlated electron-hole pairs, i.e., excitons, discussed above.
Moreover, we deal with well-precise spin selection rules: 
the only matrix elements in (\ref{H_lambda_lambdaprime}) 
different from zero are those conserving the total spin of the carrier-light 
system.
Indeed, the possible final states $\vert \lambda^{ }_{N} \rangle$ 
depend on the 
spin configuration of the initial many-body state
$\vert \lambda'_{N'} \rangle$
as well as on the polarization of the electromagnetic field $E(t)$. 
In particular, we are allowed to create two excitons with opposite spin 
orientation (i.e., antiparallel-spin configuration) in the same orbital 
quantum state.
In contrast, due to the Pauli exclusion principle, two excitons with the 
same spin orientation (i.e., parallel-spin configuration) cannot occupy the
same orbital state.

By treating Eq.~(\ref{td-Schr2}) within the standard time-dependent 
perturbation-theory approach and assuming a monochromatic light source 
of frequency $\omega$, we can define the absorption probability 
corresponding to the $\lambda^{ }_{N-1} \to \lambda^{ }_N$ transition: 
\begin{equation}\label{FGR1}
P_{\lambda^{ }_{N-1} \to \lambda^{ }_N}(\omega) = {2\pi \over \hbar} 
\left\vert H'_{\lambda^{ }_N \lambda^{ }_{N-1}} \right\vert^2
\delta({\cal E}_{\lambda^{ }_N} -{\cal E}_{\lambda^{ }_{N-1}} - \hbar\omega) 
\ .
\end{equation} 
It describes the many-exciton optical response of our QD structure, i.e., 
the probability of creating a new exciton in the presence of $N-1$ 
Coulomb-correlated electron-hole pairs.

\subsubsection{Excitonic absorption}

As a starting point, let us consider the so-called excitonic response, 
i.e., the optical response of our carrier system for the $0 \to 1$ 
transition.
In this case, the initial ($N = 0$) state is the (electron-hole) vacuum 
state $\vert 0 \rangle$, while the final ($N = 1$) state 
$\vert \lambda_1 \rangle$
corresponds to a 
Coulomb-correlated electron-hole pair, i.e., an exciton.
Combining Eqs.~(\ref{ket_lN}) and (\ref{ket_lambdaN}), for $N = 1$ we have:
\begin{equation}\label{ket_lambda1}
\vert \lambda_1 \rangle = \sum_{l_1} U^{\lambda_1}_{l_1} 
c^\dagger_{i_1} d^\dagger_{j_1} \vert 0 \rangle \ ,
\end{equation}
where $l_1=i_1,j_1$ denotes the single-particle electron-hole basis 
for $N = 1$.

The excitonic-absorption probability is then given by Eq.~(\ref{FGR1}) with
$N = 1$:
\begin{equation}\label{FGR2}
P^{ex}_{\lambda_1}(\omega) = {2\pi \over \hbar} 
\left\vert H'_{\lambda_1 0} \right\vert^2
\delta({\cal E}_{\lambda_1} - \hbar\omega) 
\ ,
\end{equation} 
where 
\begin{equation}\label{H_lambda1_0}
H'_{\lambda_1 0} = 
\langle \lambda_1 \vert {\bf H}' \vert 0 \rangle
\end{equation}
is the matrix element of the light-matter Hamiltonian (\ref{H_prime}) for 
the $0 \to 1$ optical transition. Its explicit form is given in 
App.~\ref{app:emb}.

The excitonic spectrum is finally obtained by summing the absorption 
probability in (\ref{FGR2}) over all possible final states $\vert \lambda_1
\rangle$:
\begin{equation}\label{as1}
A^{ex}(\omega) = \sum_{\lambda_1} P^{ex}_{\lambda_1}(\omega) \ .
\end{equation}

\subsubsection{Biexcitonic absorption}

Let us now come to the so-called biexcitonic response, 
i.e., the optical response corresponding to the $1 \to 2$ 
transition.
In this case, the initial ($N = 1$) state coincides with the 
excitonic state $\vert \lambda_1 \rangle$ in (\ref{ket_lambda1}), while the
final ($N = 2$) state $\vert \lambda_2 \rangle$ 
corresponds to two Coulomb-correlated electron-hole 
pairs, i.e., a biexciton.
Combining again Eqs.~(\ref{ket_lN}) and (\ref{ket_lambdaN}), for $N = 2$ 
we get:
\begin{equation}\label{ket_lambda2}
\vert \lambda_2 \rangle = \sum_{l_2} U^{\lambda_2}_{l_2} 
c^\dagger_{i_1} d^\dagger_{j_1} c^\dagger_{i_2} d^\dagger_{j_2} \vert 0 \rangle \ ,
\end{equation}
where $l_2 \equiv i_1 j_1, i_2 j_2$ 
denotes the single-particle electron-hole basis 
for $N = 2$.

The excitonic-absorption probability is then given by Eq.~(\ref{FGR1}) with
$N = 2$:
\begin{equation}\label{FGR3}
P^{biex}_{\lambda^{ }_1 \to \lambda^{ }_2}(\omega) = {2\pi \over \hbar} 
\left\vert H'_{\lambda^{ }_2 \lambda^{ }_1} \right\vert^2
\delta({\cal E}_{\lambda^{ }_2} - {\cal E}_{\lambda^{ }_1} - \hbar\omega) 
\ ,
\end{equation} 
where 
\begin{equation}\label{H_lambda2_lambda1}
H'_{\lambda^{ }_2 \lambda^{ }_1} = 
\langle \lambda^{ }_2 \vert {\bf H}' \vert \lambda^{ }_1 \rangle
\end{equation}
is the matrix element of the light-matter Hamiltonian (\ref{H_prime}) for 
the $1 \to 2$ optical transition. Its explicit form is given again in 
App.~\ref{app:emb}.

The biexcitonic spectrum is finally obtained by summing the absorption 
probability in (\ref{FGR3}) over all possible final states 
$\vert \lambda_2 \rangle$:
\begin{equation}\label{as2}
A^{biex}_{\lambda^{ }_1}(\omega) = \sum_{\lambda^{ }_2} 
P^{biex}_{\lambda^{ }_1 \to \lambda^{ }_2}(\omega) \ .
\end{equation}
We stress that, contrary to the excitonic spectrum in (\ref{as1}), 
the biexcitonic spectrum $A^{biex}$ is a function of the initial excitonic 
state 
$\lambda^{ }_1$.

\medskip

Equations (\ref{FGR2}) and (\ref{FGR3}) will be employed 
in Sect.~\ref{s:eor} to investigate the 
electro-optical response of single as well as coupled QD structures. 
However, for the case of ultrafast optical excitation and strong 
light-matter coupling, the above perturbation-theory picture can no longer
be applied, and the time evolution of our many-body state $\vert \psi(t) 
\rangle$ can be obtained by solving the time-dependent Schr\"odinger 
equation in (\ref{td-Schr1}).
We stress that, contrary to the many-exciton absorption probability in 
(\ref{FGR1}), the number of excitons, i.e.,
\begin{equation}\label{N}
N(t) 
= \langle \Psi(t) \vert {\bf N}_e \vert \Psi(t) \rangle
= \langle \Psi(t) \vert {\bf N}_h \vert \Psi(t) \rangle \ ,
\end{equation}
is a continuous function of time and changes according to the specific 
ultrafast laser-pulse sequence considered.

\subsection{Interaction with environmental degrees of freedom}\label{ss:env}

Let us finally come to the interaction of the carrier subsystem 
with various environmental degrees of freedom, like 
phonons, plasmons, etc.
They will not be treated explicitly; instead, we shall adopt a statistical 
description of the carrier subsystem in terms of its density-matrix 
operator 
\begin{equation}\label{rho1}
\rho(t) = \overline{\vert \Psi(t) \rangle \langle \Psi(t) \vert} \ ,
\end{equation}
the overbar denoting a suitable ensemble average.\cite{dma}
Its time evolution can be schematically written as
\begin{equation}\label{rho2}
\frac{d}{dt} \rho(t) 
= 
\frac{d}{dt} \rho(t)\Biggr|_{\bf H} 
+
\frac{d}{dt} \rho(t)\Biggr|_{env} \ .
\end{equation}
The first term describes the deterministic evolution induced by the carrier
Hamiltonian ${\bf H}$ according to the well-known Liouville-von Neumann 
equation
\begin{equation}\label{LvN}
\frac{d}{dt} \rho(t)\Biggr|_{\bf H} 
= {1 \over i\hbar}\, \left[{\bf H}, \rho(t)\right] \ ,
\end{equation}
while the second one describes a non-unitary evolution,\cite{PZ} 
due to energy-relaxation and dephasing processes.
The latter will be treated within the standard $T_1 T_2$ model (see 
Sect.~\ref{ss:afse}).

As for the case of the Schr\"odinger equation (\ref{td-Schr1}), it is 
convenient to describe the density-matrix operator $\rho$ ---as well as its
time evolution--- within our Coulomb-correlated $N$-pair basis.
By combining Eqs.~(\ref{Psi}) and (\ref{rho1}), we get
\begin{equation}\label{dm}
\rho_{\lambda^{ }_N \lambda'_{N'}}(t) = 
\overline{
a^{ }_{\lambda^{ }_N}(t)
a^*_{\lambda'_{N'}}(t)
} \ :
\end{equation}
the density matrix in the $\lambda$-representation is bilinear in the 
state coefficients $a_{\lambda^{ }_n}$ in (\ref{Psi}).

\subsection{The excitonic picture}\label{ss:ep}

As discussed in Sect.~\ref{ss:cccs}, the generic Coulomb-correlated 
$N$-pair state $\vert \lambda_N \rangle$ can be written as a linear 
combination [see Eq.~(\ref{ket_lambdaN})] of the non-interacting electron-hole 
basis states in (\ref{ket_lN}). Such single-particle picture is used to 
compute Coulomb-correlated $N$-pair states and energy levels via the 
exact-diagonalization approach described in App.~\ref{app:emb}.
However, it is often convenient to adopt ---in stead of a single-particle 
description--- an excitonic-like picture, i.e., a quasi-particle number 
representation based on Coulomb-coupled electron-hole pairs. 
Aim of this section is (I) to show that, in general, such an excitonic 
description is not possible, and (II) to identify the basic requirements 
needed for such a quasi-particle number representation and therefore for 
QIC processing.

To this end, let us introduce the following set of excitonic creation 
operators: 
\begin{equation}\label{X-1}
\vert \lambda_1 \rangle = X^\dagger_{\lambda_1} \vert 0 \rangle \ ,
\end{equation}
where, as usual, $\vert 0 \rangle$ denotes the electron-hole vacuum state and 
$\lambda_1$ is the label for the generic excitonic ($N = 1$) state.
By comparing Eq.~(\ref{ket_lambda1}) with the above definition, we can 
write these excitonic operators in terms of the electron and 
hole operators, i.e.,
\begin{equation}\label{X-2}
X^\dagger_{\lambda_1} = 
\sum_{ij} U^{\lambda_1}_{ij} c^\dagger_{i} d^\dagger_{j} \ .
\end{equation}
Moreover, in view of the unitary character of the transformation $U$, we 
get:
\begin{equation}\label{X-3}
c^\dagger_{i} d^\dagger_{j} = \sum_{\lambda_1} 
{U^{\lambda_1}_{ij}}^* X^\dagger_{\lambda_1} \ .
\end{equation}
If we now consider the explicit form of the non-interacting basis states 
in (\ref{ket_lN}), 
the generic $N$-pair many-body state 
(\ref{ket_lambdaN}) can formally be written as 
\begin{equation}\label{X-4}
\vert \lambda_N \rangle = \sum_{\{\lambda_1\}} 
C^{\lambda_N}_{\{\lambda_1\}} 
\vert \{\lambda_1\} \rangle
\end{equation} 
with
\begin{equation}\label{X-5}
\vert \{\lambda_1\} \rangle = 
\prod_{\lambda_1} X^\dagger_{\lambda_1} \vert 0 \rangle \ .
\end{equation}

The expansion in (\ref{X-4}) would suggest to define a sort of excitonic 
number representation in terms of the $N$-pair states 
$\vert \{\lambda_1 \} \rangle$.
We stress that, in general, this is not possible. The point is that, in
general, {\it the set of states in (\ref{X-5}) do not constitute a basis for our
$N$-pair Hilbert subspace}.
This is intimately related to the fact that ---contrary to electron and hole 
creation and destruction operators--- the excitonic operators in 
Eq.~(\ref{X-2}) do not obey canonic commutation relations.
In general, the commutator
\begin{equation}\label{X-6}
{\cal C}_{ \lambda_1, \lambda'_1}:= 
\left[ X^{ }_{\lambda_1}, X^\dagger_{\lambda'_1} \right] 
\end{equation}
is itself an operator. 
This clearly prevents the introduction of number operators and, therefore, 
of a genuine quasi-particle number representation.

As will be discussed in Sect.~\ref{s:qip},
two basic requirements are needed to perform quantum information 
processing:
(i) the tensor-product 
structure of the ``computational space'' considered, and (ii) 
the $SU(2)$ character of the raising/lowering operators acting on 
our computational subsystems, known as ``qubits''.
The main question is thus to study if ---and in which conditions--- the 
coulomb-correlated electron-hole system discussed so far may act as quantum
hardware, i.e., may be used to perform quantum information processing.
This requires to identify a set of independent degrees of freedom, qubits, 
with a $SU(2)$ character, the one of spin-${1 \over 2}$ systems.

As a starting point, one should then check if there exist a set of
independent excitonic degrees of freedom;
this corresponds to verify that for any pair of excitonic states $\lambda_1$ 
and $\lambda'_1$ the commutator of Eq. (\ref{X-6})
is equal to zero.

Let us now discuss the tensor product structure of our computational 
subspace.
To this end let us consider again the case of two qubits $a$ and $b$.
Generally speaking, we know that the Hilbert space of a bipartite system 
is ${\cal H}_a \otimes {\cal H}_b $, where  ${\cal H}_{a/b}$ are the
Hilbert spaces of the individual qubits. 
This means that if $\left\{ \vert l_a \rangle\right\}$ 
is an orthonormal basis set for ${\cal H}_{a}$ and 
$\left\{ \vert l_b \rangle \right\}$ 
is an orthonormal basis set for ${\cal H}_{b}$, 
then $\left\{ \vert l_a \rangle \otimes \vert l_b \rangle \right\}$ is a 
basis set for the whole computational space.
What one needs to test is 
the possibility of writing the many-body ground state 
---corresponding in this case to a biexcitonic state $\lambda_2$---
as the product of two independent excitonic states $\lambda^a_1$ and 
$\lambda^b_1$.
This corresponds to verify that
\begin{equation}\label{X-7}
\langle \lambda_2 \vert
\left( \vert \lambda^a_1 \rangle \vert \lambda^b_1 \rangle \right) = 1
\ .
\end{equation}

Provided that the above requirements are fulfilled, 
let us now focus on the single qubit, i.e., on one of the independent 
excitonic states $\lambda_1$. 
In this case, we want to check that the exciton creation/annihilation operators 
introduced in (\ref{X-1}) obey usual $SU(2)$ commutation relations. 
More specifically, we are interested in defining the $z$-component pseudospin 
operator $S^z_{\lambda_1}$ as
\begin{equation}\label{X-8}
S^z_{\lambda_1}
:= {1 \over 2} {\cal C}_{\lambda_1, \lambda_1}\ .
\end{equation}
In order to check that this is really a $z$-component spin operator, we 
should verify that its average value over our many-body state is either 
plus or minus one. 
Deviations from this ideal scenario can be regarded as a measure of the leakage 
from our computational space due to the presence of external, i.e., 
non-computational, 
excitonic states.

In Sect.~\ref{s:qip} we shall show that for prototypical GaAs-based 
quantum-dot molecules all the above requirements are well fulfilled and our
excitonic system can indeed be used as quantum hardware.

\section{Electro-optical response of semiconductor quantum 
dots}\label{s:eor}

In this section we shall analyze the electro-optical properties, i.e., the 
optical response in the presence of a static electric field ${\bf F}$, 
of single as well as coupled QD structures. 
While the light-matter interaction is described by the Hamiltonian 
${\bf H}'$ 
in (\ref{H_prime}), the presence of a static field ${\bf F}$ can be 
accounted for by adding to the confinement potential $V_c^{e/h}$ in 
(\ref{Schr1}) the corresponding scalar-potential term:
\begin{equation}\label{F}
V_c^{e/h}({\bf r}) \to V_c^{e/h}({\bf r}) \pm e {\bf F \cdot r} \ .
\end{equation}
Here, the $\pm$ sign refers, respectively, to electrons and holes. 
As discussed below, this sign difference will give rise to exciton-exciton 
coupling and significant field-induced energy renormalizations.

Within the usual envelope-function picture,\cite{Bastard}
the single-particle properties of our quasi-0D
structure are described by the Schr\"odinger equation in (\ref{Schr1}).
Similar to the case of semiconductor quantum wires,\cite{QWR} a quantitative 
analysis of the whole single-particle spectrum requires a direct numerical 
solution of Eq.~(\ref{Schr1}); this can be performed using a 
fully 3D plane-wave expansion, as described in Ref.~\onlinecite{Troiani}  and 
briefly recalled in App.~\ref{app:spp}.

If, in contrast, our interest is limited to the low-energy range only, 
for most of the QD structures realized so far the carrier confinement 
can be described as the sum of two potential profiles, one acting along the
growth (or perpendicular) direction and one affecting the in-plane 
(or parallel) coordinates 
only:
\begin{equation}\label{V_c}
V_c^{e/h}({\bf r}) = V^{e/h}_\perp(r_\perp) + 
V^{e/h}_\parallel({\bf r}_\parallel) \ .
\end{equation} 
As a consequence, the 3D carrier envelope function $\psi_{i/j}$ can be 
factorized according to:
\begin{equation}\label{perpar1}
\psi_{i/j}({\bf r}) = \psi^\perp_{i_\perp/j_\perp}(r_\perp) 
\psi^\parallel_{i_\parallel/j_\parallel}({\bf r}_\parallel) 
\end{equation}
and the single-particle spectrum is the sum of the 
parallel and perpendicular ones:
\begin{equation}\label{perpar2}
\epsilon_{i/j} = \epsilon^\perp_{i_\perp/j_\perp} + 
\epsilon^\parallel_{i_\parallel/j_\parallel} \ .
\end{equation}
In this case, the original 3D problem is reduced to the solution of two 
independent Schr\"odinger equations, along the growth direction and within the
parallel plane. This can be still performed employing the plane-wave-expansion 
approach described in App.~\ref{app:spp}.

For most of the state-of-the-art QD structures we have strong 
confinement (few nanometers) along the growth direction, while the in-plane 
confinement potential $V^{e/h}_\parallel$ 
is much weaker. 
Moreover, as far as the low-energy 
region is concerned, the in-plane confinement is well described in terms of
a 2D parabolic potential. 
For this case the Schr\"odinger equation within the 2D parallel subspace
can be solved analytically ---also in the presence of the static field 
${\bf F}$ (see App.~\ref{app:spp})---
and thus the problem reduces to a numerical 
solution of the Schr\"odinger equation along the perpendicular direction.

The analysis of the electro-optical response of semiconductor quantum dots 
presented in the reminder of this section is based on this 
parabolic-confinement model, whose derivation and validity limits are 
discussed in App.~\ref{app:spp}.

\subsection{Single QD structure}\label{ss:sqd}

Let us start our analysis by considering the case of a single QD structure 
in the presence of an in-plane static electric field: 
${\bf F} = (F_\parallel, F_\perp = F, 0)$.
Within the parabolic-confinement model previously introduced (see also 
App.~\ref{app:spp}), the quasi 0D carrier confinement for both electrons 
and holes is described by an in-plane parabolic potential 
$V^{e/h}_\parallel$ [see Eq.~(\ref{A:pcm1})] plus a square-like potential 
$V^{e/h}_\perp$ corresponding to the interface band offset along the 
growth direction. 

As a starting point, we have considered an ideal QD structure characterized
by the following material and confinement parameters for electrons and 
holes: effective mass $m_e = 0.05 m_\circ$, $m_h = 0.08 m_\circ$ and  
parabolic-confinement energy 
$\hbar\omega_e = 40$\,meV,
$\hbar\omega_h = 25$\,meV;
The well width is $w = 50$\,\AA\  
and the static dielectric constant [see Eq.~(\ref{B:cme3})] 
has been taken to be 
$\varepsilon_0 = 12$.
Within this ideal QD model, the square-like potential profile along the growth 
direction is characterized by an infinite barrier height, i.e.
$\Delta_{e/h} \gg {\hbar^2\pi^2 \over 2 m_{e/h} w^2}$. 
By choosing the above material and confinement parameters for electrons and 
holes, we deal with a very special case for which the set of 
electron and hole single-particle envelope functions $\psi^e$ and $\psi^h$ 
coincide. Indeed, the in-plane spatial extension $\alpha$ in 
Eq.~(\ref{A:alpha})  is the same for electrons and holes.
Moreover, we shall discuss how this symmetry, not present in a realistic QD
structure (see below), is related to special features in the optical 
response of the system (``hidden symmetry''), as described in 
Ref.~\onlinecite{Hawry}.

Due to the strong perpendicular carrier confinement, for both electrons and 
holes we deal with a single localized state; 
therefore, the low-energy single-particle spectrum 
is simply given by a sequence of equally spaced discrete levels 
corresponding to the 2D parabolic confinement [see Eq.~(\ref{A:epsilon_par})].
This scenario is not affected by the presence of the in-plane static field 
$F$, which manifests itself only through an overall red-shift 
$\Delta{\cal E}$ of the single-particle spectrum, known as Stark shift 
[see Eq.~(\ref{A:Deltaepsilon})].

In the absence of Coulomb interaction, both the excitonic and the 
biexcitonic absorption spectra [see Eqs.~(\ref{as1}) and (\ref{as2})] 
will exhibit optical transitions connecting the above single-particle 
energy levels. As usual, their amplitude is dictated by the corresponding 
optical matrix elements, i.e., oscillator strength, as well as by the 
combined state degeneracy, i.e., joint density-of-states (DOS).

Figures \ref{f:sqd-abs1} and \ref{f:sqd-abs2}(B) show 
the excitonic-absorption spectrum for 
$F = 0$ and $F = 50$\,kV/cm, respectively.
Moreover, the excitonic-absorption spectra are compared 
to the single-particle ones, i.e.,the ones evaluated in the absence of Coulomb 
interaction,which respectively correspond to the dashed curves in figure 
\ref{f:sqd-abs1} and to part (A) in figure \ref{f:sqd-abs2}. 
As already pointed out, in this case the optical transitions connect the 
equally-spaced single-particle electron and hole states.

As discussed in App.~\ref{app:spp}, for $F = 0$ the only allowed
optical transitions are those conserving the envelope function
total angular momentum, i.e.,
$m = -m'$ [see Eq.~(\ref{A:amc})]; moreover, due to the special symmetry 
between electrons and holes previously discussed, we have 
$n_\epsilon = n'_\epsilon$ [see Eq.~(\ref{A:epsilon_par})].
Their amplitude is dictated by the joint
state degeneracy, which for the single-particle case (see dashed line) 
is given by $(n_\epsilon + 1)$.
In contrast, for finite values of the in-plane static field $F$ (see
part (A) in Fig.~\ref{f:sqd-abs2}), the above 
selection rules are relaxed (see App.~\ref{app:spp}) and we deal with 
new optical transitions corresponding to $m + m' \neq 0$ and 
$n_\epsilon \neq n'_\epsilon$, 
not present in the field-free case. 
Moreover, in the presence of the static field 
the spectra exhibit a significant reduction
in oscillator strength. This is ascribed to a reduction of the 
in-plane overlap between electrons and holes [see Eq.~(\ref{A:mupar2})], 
due to the charge separation induced by the applied field [see 
Eq.~(\ref{A:d_eh})].
This can be clearly seen in Fig.~\ref{f:sqd-cd1b}, where we show the 
single-particle electron and hole ground-state charge distributions (dashed
curves) corresponding to the single-particle spectra of 
Fig.~\ref{f:sqd-abs2}.
For $F = 0$ (see Fig.~\ref{f:sqd-cd1a}) the electron and hole 
parabolic-potential minima 
coincide and, therefore, the two charge distributions exhibit the same 
symmetry center.
In contrast, in the presence of the in-plane field $F$ the two potential 
minima are shifted toward different directions. 
This, in turn, induces an 
electron-hole charge separation, as clearly shown in Fig.~\ref{f:sqd-cd1b}.
Such charge displacement ---which corresponds to the formation of an in-plane
electrical dipole--- is responsible for the oscillator-strength reduction in 
Fig.~\ref{f:sqd-abs2} previously discussed.

Let us now come to the Coulomb-correlated case 
(see solid curves in Figs.~\ref{f:sqd-abs1}, \ref{f:sqd-abs2} 
and  Figs.~\ref{f:sqd-cd1a}, \ref{f:sqd-cd1b}). 
In the presence of Coulomb interaction ---which for the excitonic case 
($N = 1$) corresponds to electron-hole attraction--- 
the main effect is a global red-shift of the Coulomb-correlated spectrum 
compared to the single-particle one.
More precisely, for $F = 0$ we find a relatively strong red-shift
of the lowest optical transition (of about $20$\,meV). 
For higher transitions this effect is reduced, which can be understood 
considering that high-energy states are characterized by an increasing spatial 
extension and, therefore, by a larger average distance between 
electrons and holes.
Moreover, the Coulomb-correlated spectrum exhibits a 
transfer of oscillator strength toward low energies between 
quasi-degenerate optical transitions. 
This scenario is well established, and characterizes also systems of higher 
dimensionality, like quantum wells and wires.\cite{SST,QWR}
For increasing values of the applied field we have a progressive reduction 
of the excitonic red-shift as well as of the oscillator-strength transfer, 
i.e., of the electron-hole attraction. This is confirmed by the inset of 
Fig.~\ref{f:sqd-abs1}, where the exciton binding energy is reported as a 
function of the applied field.

In order to better understand the physical origin of this field-dependent 
behavior, we have carried on a detailed investigation of the excitonic 
wavefunction projected into the electron and hole subspaces.
More precisely, by rewriting Eq.~(\ref{ket_lambda1}) in the coordinate 
representation, the two-body excitonic wave function is given by:
\begin{equation}\label{Psiex}
\Psi^{ex}_\lambda({\bf r}_e,{\bf r}_h) = 
\sum_{ij} U^\lambda_{ij} \psi_i({\bf r}_e) \psi_j({\bf r}_h) \ .
\end{equation}
The square modulus of $\Psi^{ex}$ will then describe the conditional probability
of finding the electron with coordinate ${\bf r}_e$ and the hole with 
coordinate ${\bf r}_h$. If we now integrate such quantity over one of the 
two coordinates we get
\begin{equation}\label{psie}
f^e_\lambda({\bf r}_e) = 
\int \left\vert \Psi^{ex}_\lambda({\bf r}_e,{\bf r}_h)\right\vert^2
d{\bf r}_h
= \sum_{ii',j} U^{\lambda *}_{ij} U^\lambda_{i'j} 
\psi^*_i({\bf r}_e) \psi^{ }_{i'}({\bf r}_e)
\end{equation}
and 
\begin{equation}\label{psih}
f^h_\lambda({\bf r}_h) = 
\int \left\vert \Psi^{ex}_\lambda({\bf r}_e,{\bf r}_h)\right\vert^2
d{\bf r}_e
= \sum_{i,jj'} U^{\lambda *}_{ij} U^\lambda_{ij'} 
\psi^*_j({\bf r}_h) \psi^{ }_{j'}({\bf r}_h) \ .
\end{equation}
The quantity $f^{e/h}_\lambda$
 can be regarded as an effective single-particle 
probability distribution, which 
accounts for the electron-hole correlation described by the excitonic wave 
function in Eq.~(\ref{Psiex}). 
In the absence of Coulomb correlation the transformation $U$ reduces to the
identity ($U^\lambda_{ij} = \delta_{\lambda, ij}$) and the effective 
single-particle distributions $f^{e/h}_\lambda$ 
coincide with the square modulus of
the single-particle wavefunctions of electrons and holes, i.e.,
$f^e_i({\bf r}_e) = \vert\psi_i({\bf r}_e)\vert^2,
f^h_j({\bf r}_h) = \vert\psi_j({\bf r}_h)\vert^2$.

The effective charge distributions for electrons and holes --defined, 
respectively, in (\ref{psie}) and (\ref{psih})---
are plotted in Fig.~\ref{f:sqd-cd1a} and in Fig.~\ref{f:sqd-cd1b} 
for the ground-state-exciton case.
As expected, in the presence of Coulomb correlation the charge distribution
deviates from the corresponding Coulomb-free case (dashed 
curves).
For $F = 0$ (Fig.~\ref{f:sqd-cd1a}) the average distance 
between electrons and holes 
is very limited, which leads to a strong exciton binding (see 
Fig.~\ref{f:sqd-abs1}).
For increasing values of the applied field (Fig.~\ref{f:sqd-cd1b})
we see again an increasing charge separation. However, the effect is 
now reduced, compared to the Coulomb-free case (see dashed curves). 
This is due to the competition between the displaced parabolic 
potentials  and the electron-hole Coulomb attraction (see also 
Fig.~\ref{f:Veff} in Sect.~\ref{ss:sm}). 
The latter is 
progressively reduced due to a significant increase of the electron-hole 
average distance (see again Fig.~\ref{f:sqd-cd1b}). 
This also explains the reduction of the excitonic binding energy reported as 
inset in Fig.~\ref{f:sqd-abs1}.

The analysis presented so far 
suggests that the behavior of the system is governed by the following three 
characteristic lengths:
\begin{itemize}
\item[(i) ] 
the radial extension $\alpha$ of the
parabolic ground state, which in this case is the same for 
electrons and holes [see Eq.~(\ref{A:alpha})]; 
\item[(ii) ]
the excitonic Bohr 
radius 
\begin{equation}\label{aex}
a^{ex} =  \frac{\hbar^2\varepsilon_0}{e^2\mu}  \ ,
\end{equation}
where $\mu$ is the reduced electron-hole mass;
\item[(iii) ]
the total electron-hole displacement 
$ d =\vert {\bf d}^h_\parallel - {\bf d}^e_\parallel \vert$ 
[see Eq.~(\ref{A:d_eh})].
\end{itemize}
\noindent
Generally speaking, when $\alpha \ll a^{ex}$ we are in the so-called 
strong-confinement limit: the carrier confinement is dictated by the 
single-particle parabolic potential only, which implies that the wavefunction 
of the excitonic ground state coincides with the product of the electron 
and hole single-particle wavefunctions, i.e., the expansion 
in Eq.~(\ref{Psiex}) contains just one term.
In the opposite case, called weak-confinement limit ($\alpha \gg a^{ex}$), 
the excitonic wavefunction depends on the relative coordinate only and 
resembles the 2D hydrogen-atom solution.

For the case under investigation the situation is as follows.
In the field-free case ($d = 0$), the excitonic Bohr radius 
($a^{ex} \simeq 200$\,\AA) is of the same order of the electron and hole 
ground-state radial extension ($\alpha \simeq 60$\,\AA), 
which implies that the exciton wavefunction deviates from the product of 
the corresponding single-particle states. 
This is confirmed by the 
Coulomb-correlated carrier distribution of 
Fig.~\ref{f:sqd-cd1a}, compared to the Coulomb-free one 
(dashed curve).
However, we are not very far from the ideal strong-confinement limit 
previously discussed; Indeed, our numerical analysis has shown that the 
single-particle expansion in (\ref{Psiex}) can be limited to 
a relatively small number ($6 \times 6$) of electron-hole states.
For increasing values of the applied field ---and, therefore, of the 
charge displacement $d$--- the average distance between electrons and holes
increases, thus reducing Coulomb-correlation effects.
This is confirmed by the absorption spectra in Fig.~\ref{f:sqd-abs2} as 
well as by the carrier distributions in Fig.~\ref{f:sqd-cd1b}, where the 
difference
between Coulomb-correlated and Coulomb-free results is significantly reduced. 
We can therefore conclude that the presence of an in-plane static field 
$F$
induces a net electron-hole charge separation, which leads to a 
significant suppression of electron-hole Coulomb correlation.

Let us now move to the biexcitonic response of our ideal QD system. 
As discussed in Sect.~\ref{ss:cl}, contrary to the excitonic case 
investigated so far, the latter depends on the spin configuration of both 
initial ($N = 1$) and final ($N = 2$) Coulomb-correlated states.
More precisely, due to the spin selection rules in the light-matter 
interaction Hamiltonian, we deal with two relevant cases only: the 
parallel- and the antiparallel-spin one.

Let us consider first the antiparallel-spin configuration. In this case both 
excitons can occupy the low-energy orbital state.
Figure \ref{f:sqd-abs3} shows the biexcitonic spectrum (dashed curve) 
compared to the excitonic one (solid curve) for $F = 30$\,kV/cm.
We can clearly identify a biexcitonic transition (see first 
peak of the dashed curve), which is blue-shifted with respect to the 
ground-state excitonic transition (see first peak of the solid curve). 
This energy 
renormalization is known as {\it biexcitonic shift}:\cite{QDs}
\begin{equation}\label{DeltacalE1}
\Delta{\cal E} = 
{\cal E}_{\lambda_2} - {\cal E}_{\lambda_1} - {\cal E}_{\lambda'_1} \ .
\end{equation}
This positive energy shift can be understood as follows: 
the applied field induces for 
both excitons the same  
charge separation 
(see Fig.~\ref{f:sqd-cd1b}) 
which results in a repulsive dipole-dipole coupling.
This is confirmed by the field-dependent behavior of the biexcitonic 
splitting $\Delta{\cal E}$ reported as inset in Fig.~\ref{f:sqd-abs3}.

Moreover, as shown in the inset, in the field-free case,  
the dot behaves as an artificial atom and the
energy to add an exciton in a shell is, up to the first order in the Coulomb 
interaction, independent of the shell 
occupation.\cite{HawryPRB} 
Indeed, within first-order perturbation theory, when $F=0$ the two 
excitons occupy the same 
spherically symmetric orbital ground state and 
for an ideal structure as the one we are considering, the 
biexcitonic splitting is exactly zero, 
because in this case the various attractive 
and 
repulsive Coulomb interactions cancel exactly.\cite{Hawry}
This can be understood as follows.
In the strong-confinement limit ---which is not far from
the regime considered here--- and for antiparallel spins, the biexcitonic 
splitting can be very  well approximated by the non-interacting
single-particle probability 
distributions 
$f^e_0 = \vert \psi^e_0 \vert^2$ and $f^h_0 = \vert \psi^h_0 \vert^2$ 
only, i.e., 
\begin{equation}\label{DeltacalE2}
\Delta{\cal E} = {e^2 \over \varepsilon_0}\, \int d{\bf r} \int d{\bf r}' 
{ \Delta f({\bf r}) \Delta f({\bf r}') 
\over
\vert {\bf r-r'}\vert } \ ,
\end{equation}
where $\Delta f = f^e_0 - f^h_0$ is the difference between the electron and 
hole single-particle probability densities. 
Due to the special symmetry of the ideal QD structure under investigation, 
in the field-free case we have $\Delta f = 0$ and  the biexcitonic 
splitting is zero as well. 

As already pointed out, in the QD structure under investigation
we are not far from the strong 
confinement limit. However,
since our calculation of the biexcitonic splitting is non-perturbative, 
we get,
even in the field-free case, a non-vanishing  $\Delta{\cal E}$.
This small, but not negligible biexcitonic splitting 
($\Delta{\cal E}(F=0) = 0.7$\,meV, see inset in Fig.~\ref{f:sqd-abs3})
measures the Coulomb-interaction contribution, underlying that the real 
ground-state 
biexcitonic wave function has contributions also from higher-level 
single-particle states. 
This  value is
compatible with the one given in Ref.~\onlinecite{QDs}. \\
We shall now show that the above field-free behavior is due to the special
choice of material and confinement parameters of the ideal QD structure 
investigated so far.
To this aim, let us now move to the case of a realistic semiconductor macroatom. 
As prototypical system let us consider a GaAs/AlAs QD structure 
characterized by the following material parameters:
effective masses $m_e = 0.067 m_\circ$ and $m_h = 0.34 m_\circ$, 
conduction- and valence-band offsets $\Delta_e = 1$\,eV and $\Delta_h = 
0.58$\,eV, 
parabolic-confinement energies $\hbar\omega_e = 30$\,meV and $\hbar\omega_h
= 24$\,meV, 
well width $w = 50$\,\AA, and static dielectric constant 
$\varepsilon_0 = 12.1$.

Figure \ref{f:sqd-abs4} shows again the comparison between excitonic 
(solid curve) and biexcitonic spectrum (dashed curve) 
for the antiparallel case in the 
field-free case.
As we can see, contrary to the result in Fig.~\ref{f:sqd-abs3}, we now deal
with a significant biexcitonic splitting $\Delta{\cal E}$ also 
in the absence of the 
in-plane field (see inset in Fig.~\ref{f:sqd-abs4}).
Indeed, for any realistic QD structure we deal with different spatial 
extensions $\alpha_e$ and $\alpha_h$ of the electron and hole single-particle 
in-plane ground states. 
In Fig.~\ref{f:sqd-cd2} we report the electron and hole single-particle charge 
distributions $f^e$ and $f^h$ (solid curves) as well as their difference 
$\Delta f$ (dashed curve).
As anticipated, due to the different material and confinement parameters, 
the charge distributions for electrons and holes do not coincide anymore.
This, in turn, gives rise to local violations of charge neutrality, i.e., 
$\Delta f \neq 0$, and therefore to a non-vanishing biexcitonic shift [see 
Eq.~(\ref{DeltacalE2})].
We finally stress that the presence of the in-plane static field leads to a
further increase of $\Delta {\cal E}$ (see inset in 
Fig.~\ref{f:sqd-abs4}).

Let us now move to the parallel-spin configuration. 
In this case we study the probability of creating a 
second exciton in the dot with the same spin orientation of the first one. 
Due to the Pauli exclusion principle, the two excitons are not allowed to 
occupy the same exciton state. 
As already pointed out [see Eq.~(\ref{as2})], 
the biexcitonic spectrum of
the system depends on its initial excitonic state 
$\vert \lambda_1 \rangle$.
In Fig.~\ref{f:sqd-abs5} we compare the biexcitonic spectrum (dashed 
curve) with the corresponding excitonic spectrum (solid curve) for the 
field-free case. Here, the biexcitonic spectrum has been computed 
assuming, as initial state $\vert \lambda_1 \rangle$, the excitonic ground 
state. 
Let us focus on the low-energy part of the spectrum: as expected, due to 
the Pauli principle, the exciton ground state is forbidden to the second 
exciton, which can occupy any other high-energy state. 
Contrary to the antiparallel case (see inset in Fig.~\ref{f:sqd-abs4}), 
we now deal with a negative biexcitonic
shift $\Delta{\cal E}$, i.e., the lowest biexcitonic transition (solid 
curve) is red-shifted compared to the corresponding 
excitonic one (second peak of the dashed curve). 
As discussed in Ref.~\onlinecite{Troiani}, such energy renormalization 
(in this case of the order of $10$\,meV) can be 
ascribed to the attractive dipole-dipole interaction between the two 
excitons in the dot. Indeed, due to their different orbital quantum 
numbers $\lambda_1$ and $\lambda'_1$ ---and thus to their different spatial 
charge distributions--- they exhibit significant Coulomb coupling.
Contrary to the antiparallel-spin case previously discussed, 
(see inset in Fig.~\ref{f:sqd-abs4}), 
now the presence of an in-plane static field $F$ 
leads to a reduction and eventually to an inversion of the biexcitonic 
shift (see inset in 
Fig.~\ref{f:sqd-abs5}).
This can be understood as follows: the application of the in-plane field 
leads to a progressive reduction of the attractive, i.e., spatially 
antiparallel, dipole-dipole Coulomb coupling; for high fields this is 
transformed into a repulsive, i.e., spatially parallel, dipole-dipole 
interaction and therefore to a positive biexcitonic splitting.
The transition from  red biexcitonic shifts to blue ones, occurs when 
the displacement induced by the electric field becomes of the same order or 
bigger than the excitonic Bohr radius. 

\subsection{Coupled QD structure}\label{ss:cqd}

Let us now consider the case of a semiconductor macromolecule, i.e., a 
coupled QD structure.
In particular, as prototypical system we shall investigate the GaAs/AlAs 
coupled QD structure schematically depicted in Fig.~\ref{f:cqd-str}. The 
material and confinement parameters are the same of the realistic single-QD
structure previously investigated (see Figs.~\ref{f:sqd-abs4}, 
\ref{f:sqd-abs5}, and \ref{f:sqd-cd2}):
effective masses $m_e = 0.067 m_\circ$ and $m_h = 0.34 m_\circ$, 
conduction- and valence-band offsets $\Delta_e = 1$\,eV and $\Delta_h = 
0.58$\,eV, 
parabolic-confinement energies $\hbar\omega_e = 30$\,meV and $\hbar\omega_h
= 24$\,meV, 
static dielectric constant $\varepsilon_0 = 12.1$.
The square-like carrier confinement along the growth direction for 
electrons and holes is schematically depicted in Fig.~\ref{f:cqd-str} for 
our semiconductor macromolecule $a+b$. 
This is tailored in such a way to allow for
an energy-selective creation/destruction of bound electron-hole pairs in 
dots $a$ and $b$.
Indeed, the width of wells $a$ and $b$ are slightly different, which 
corresponds to a blue-shift of about $10$\,meV of the single-particle 
optical transitions of dot $b$ with respect to the corresponding transition 
in dot $a$.
We stress that such energy shift is also present in the absence of 
inter-dot tunneling and Coulomb coupling.
Moreover, the inter-dot barrier 
width ($w$ $\sim$ $50$\,\AA) is such to prevent
single-particle tunneling and at the same
time to allow for significant inter-dot Coulomb coupling.
We stress that the geometrical and material parameters of the proposed 
prototypical structure in Fig.~\ref{f:cqd-str} are fully compatible 
with current
QD growth and characterization technology \cite{QDs,CCC}.

Let us discuss first the excitonic response of the 
{\it semiconductor macromolecule} ($a+b$) in Fig.~\ref{f:cqd-str}. 
The excitonic ($0 \to 1$) optical spectrum in the presence of an in-plane 
electric field $F = 75$\,kV/cm is shown in 
Fig.~\ref{f:cqd-abs1}. Here, the Coulomb-correlated result (B) is
compared to the Coulomb-free one (A). The scenario is very 
similar to the single-dot case previously investigated (see 
Fig.~\ref{f:sqd-abs1}): for relatively strong values of the applied field, 
apart from a rigid red-shift, the Coulomb-correlated result is very similar
to the Coulomb-free one.
Here, the two lowest optical transitions 
correspond to the formation of direct ground-state excitons in dot $a$ and 
$b$, respectively (see Fig.~\ref{f:cqd-str}). 
In contrast, the high-energy peaks correspond to optical transitions 
involving excited states of the in-plane parabolic potential.
Due to the strong in-plane carrier confinement ---compared to the 
relatively large electron-hole charge displacement---
the two low-energy excitonic 
states are expected to closely resemble the 
corresponding single-particle ones. 

Let us now come to the biexcitonic response of our semiconductor 
macromolecule. In view of the strong-confinement regime considered, we shall 
focus on the two ground-state excitons only. Moreover, since we are 
primarily interested in studying inter-dot Coulomb coupling, we shall 
consider the parallel-spin configuration. 

In Fig.~\ref{f:cqd-abs2} the excitonic spectrum (solid curve) is compared
to the biexcitonic one (dashed curve).
The latter describes the generation of a second electron-hole 
pair in the presence of a previously created exciton ($1 \to 2$ 
optical transitions). In particular, here the previously generated exciton 
is assumed to be in dot $a$.
As for the single-dot case previously investigated (see 
Fig.~\ref{f:sqd-abs4}), the crucial feature in Fig.~\ref{f:cqd-abs2} is the 
magnitude of the biexcitonic shift.
For the QD structure under investigation we get energy splittings up 
to $8$\,meV (see inset in Fig.~\ref{f:cqd-abs2}). 
This can be ascribed again to the in-plane static field $F$, 
which induces, in both dots, the excitonic dipole previously investigated 
(see Fig.~\ref{f:sqd-cd1b}).
This, in turn, gives rise to significant inter-dot dipole-dipole
coupling between adjacent excitonic states. 
The microscopic nature of such exciton-exciton coupling is 
the same of the Forster process exploited by Quiroga and Johnson~\cite{Q-J}
for the generation of entangled states in coupled QDs.

The physical origin of the biexcitonic shift $\Delta {\cal E}$ 
is qualitatively described in
Fig.~\ref{f:cqd-str}, 
where we show the effective spatial charge distribution of the two 
electrons ($e_a$ and $e_b$) and holes ($h_a$ and $h_b$) 
corresponding to the biexcitonic ground state in Fig.~\ref{f:cqd-abs2}.
As we can see, the charge separation
induced by the static field increases significantly the average distance
between electrons and holes, thus decreasing their attractive interaction.
On the other hand, the repulsive terms are basically field independent.
This is the origin of the positive energy difference  $\Delta {\cal E} $  in
Fig.~\ref{f:cqd-abs2}. 

Let us now investigate the possibility of using such QD molecules as quantum 
hardware for QIC processing. As discussed in Sect.~\ref{ss:ep}, to this end 
a few basic requirements should be fulfilled. 
First of all, the operators for the two ground-state excitons in 
dots $a$ and $b$ should commute. By evaluating the average value (over the 
biexcitonic ground state) of the 
commutator in Eq.~(\ref{X-6}), this came out to be negligibly small, thus 
confirming that these are indeed independent degrees of freedom.
Moreover, 
due to the relatively large inter-dot distance ---compared to the 
spatial extension of the carrier wavefunctions along the growth direction---
the biexcitonic ground state in Fig.~\ref{f:cqd-str} is expected to closely
resemble the product of the two excitonic states in dots $a$ and $b$.
Indeed, for the coupled QD structure under investigation we find that the 
scalar product in Eq.~(\ref{X-7}) gives a value of $0.99$, very close to 
1.
The product structure for the bipartite system 
Hilbert space is therefore very well achieved. 
It is worth noticing that, in the case in which the two excitonic states  
are localized on the same dot, e.g., in the ground and first excited 
states,
one gets a smaller value of about $0.9$.
This is a clear indication that the tensor product structure for the 
many-body state is much better achieved in a coupled QD structure than in a 
single QD system, as the one proposed in Ref.~\onlinecite{Troiani}.

Let us finally focus on the $SU(2)$ character of our excitonic qubits.
To this end, we have evaluated the average value (over the biexcitonic 
ground state) of the pseudospin operator $S^z$ introduced in Eq.~(\ref{X-8}).
By truncating the single-particle basis considering just the lowest 
energy level in each QD (strong-confinement limit),
one gets 
$\langle \lambda_2 \vert S^z_{\lambda_1} \vert \lambda_2 \rangle = 1$,
thus confirming that the operators in Eqs.~(\ref{X-6}) and (\ref{X-8}) 
are the generators of a $SU(2)$ algebra. 
In contrast, far from the strong-confinement limit,
we get a result which is of course dependent on how many 
single-particle states contribute to form an exciton.
Therefore, if we calculate again the 
mean value of the commutator considering an enlarged single-particle basis set,
we get deviations from the above ideal result.
As anticipated in Sect.~\ref{ss:ep}, 
this turns out to be a measure of the leakage of our qubit.

We can therefore conclude that ground-state excitonic transitions in our 
coupled QD molecule fulfill all the basic requirements for a 
semiconductor-based implementation of QIC processing. They can be used as 
computational degrees of freedom, i.e., qubits, and the standard pseudospin 
language can be employed.

\subsection{A simplified model}\label{ss:sm}

In this section we shall present a simplified model able
to properly describe excitonic binding as well as interdot biexcitonic 
coupling. 
Its analytical solution will allow for an extremely quick way of 
identifying suitable parameter sets 
needed to employ the above coupled QD structure as semiconductor-based
hardware for QIC processing (see Sect.~\ref{s:qip}).

As a starting point, let us consider again the typical single-QD 
structures of Sect.~\ref{ss:sqd}, 
whose single-particle confinement is modeled in terms of a box-like 
potential of width $a$ in the growth (or perpendicular) direction 
and a 2D parabolic 
potential in the in-plane (or parallel) directions.
As previously discussed, this allows a factorization of the original 3D 
single-particle problem into a perpendicular and a parallel one [see 
Eqs.~(\ref{perpar1}) and (\ref{perpar2})]. However, in the presence of Coulomb 
interaction such factorization is, in principle, no-longer possible.

More specifically, let us consider the single-exciton problem ($N = 1$) 
in the presence of an in-plane electric field ${\bf F}$;
this can be described by the  two-body Hamiltonian
\begin{equation}\label{twobody1}
{\bf H}={\bf H}^e({\bf r}_e) + {\bf H}^h({\bf r}_h) 
-{e^2 \over \varepsilon_0 \vert {\bf r}_e -{\bf r}_h \vert}\ .
\end{equation}
As discussed in App.~\ref{app:spp}, the single-particle Hamiltonians for 
electrons and holes can be written in the compact form 
[see Eqs.~(\ref{A:Schr2}), (\ref{A:Schr3}), and (\ref{A:Schr4})]:
\begin{equation}\label{sm:2body2}
{\bf H}^{e/h}({\bf r}) = -{\hbar^2\nabla^2_{{\bf r}} \over 2 m_{e/h} }
+ V^{e/h}_\perp(r_\perp) 
+ {1 \over 2} m_{e/h} 
\omega_{e/h}^2 \vert {\bf r}_\parallel - {\bf d}^{e/h}_\parallel\vert^2
+ \Delta\epsilon_{e/h} \ .
\end{equation}
Here, the presence of the applied field results in a displacement  
${\bf d}^{e/h}_\parallel$ [see Eq.~(\ref{A:d_eh})] 
of the parabolic-potential minimum as well as in a rigid energy shift 
$\Delta\epsilon_{e/h}$ 
[see Eq.~(\ref{A:Deltaepsilon})].
We want to show that for all the QD structures previously investigated
Eq.~(\ref{twobody1}) can be approximated to an analytically solvable form,
 and important quantities as
wave functions or biexcitonic shifts can be easily estimated  with a good
degree of accuracy.

In our QD structures the
wavefunction is strongly confined along the growth direction by the square well
potential, so that we can approximate 
$(r^e_\perp - r^h_\perp)^2$ 
in the Coulomb term with its average value $l^2$. 
We choose $l$ to be  twice
the
average length related to the ground state of an {\it infinite-height} square
well 
of width $a$, i.e.,
$l
= (2a/\pi)\sqrt{(\pi^2-6)/12}$.
It is thus possible to separate the Hamiltonian (\ref{twobody1}) as
${\bf H} = {\bf H}_\parallel({\bf r}_{e\parallel},{\bf r}_{h\parallel})
 + {\bf H}_\perp(r_{e\perp}) + 
{\bf H}_\perp(r_{h\perp})$,
where
${\bf H}_\perp(r_{i\perp}) = {p_{r_{i\perp}}^2/ 2m_i} +V^i_c(r_{i\perp})$
is the single-particle Hamiltonian along the growth direction 
---exactly solvable for the case of a parabolic potential as well as of 
an infinite-height  square well.
By further defining the center of mass (CM) and  relative coordinates 
${\bf R} = [m_e ({\bf r}_{e\parallel}+{\bf d}^e_{\parallel}) + m_h 
({\bf r}_{h\parallel}-{\bf d}^h_{\parallel})]/ M$, 
($M = m_e + m_h$)
and
${\bf r} = {\bf r}_{h\parallel} - {\bf r}_{e\parallel}$, 
the in-plane Hamiltonian ${\bf H}_\parallel$ becomes
\begin{eqnarray}
{\bf H}_\parallel({\bf R},{\bf r}) 
&=& 
{P^2\over 2M}  + {1\over 2} M\omega_R^2 R^2 +
 {p\over 2\mu } + {1\over 2} \mu\omega_r^2 |{\bf d}-{\bf r} |^2 
 \nonumber \\ 
& & + \mu (\omega_e^2 - \omega_h^2) 
{\bf R}\cdot ({\bf d}-{\bf r})
- {e^2\over\varepsilon_0 \sqrt{r^2+l^2}} \ ,
\label{parH2}
\end{eqnarray}
where $\mu = {m_em_h / M}$ is the reduced mass,
$\omega_R^2 = (1 +\Delta)(\omega_e^2 + \omega_h^2)/ 2$,
$\omega_r^2 = (1 -\Delta)(\omega_e^2 + \omega_h^2)/  2$,  
$\Delta = [(m_e-m_h) / M] 
(\omega_e^2-\omega_h^2) / (\omega_e^2+\omega_h^2)$ and 
\begin{equation}\label{d}
{\bf d} = -{\bf d}^e_\parallel+{\bf d}^h_\parallel  
= e{\bf F}({1\over
m_e\omega_e^2}+{1\over m_h\omega_h^2})
\end{equation}
 denotes the total (electron+hole) field-induced in-plane displacement.
 
In the limit 
$(\omega_e^2-\omega_h^2)/  (\omega_e^2+\omega_h^2) \ll 1$, 
the two coordinates are only weakly coupled, and 
the Schr\"odinger equation associated to the CM coordinate ${\bf R}$ is 
exactly solvable;
in the general case, we shall concentrate on the ground state, though 
the generalization to higher states is straightforward.
We  can approximate the ground state of 
${\bf H}_\parallel$ as
$
\Psi({\bf r},{\bf R}) =
\Psi_x(x) \chi(y,{\bf R})$,
where $x$ and $y$ denote, respectively, the components of ${\bf r}$ 
parallel and perpendicular to the field ${\bf F}$, $\chi(y,{\bf R}) =
e^{-{y^2 \over 2 \lambda_r^2}}/(\lambda_r^2\pi)^{1/4}
\cdot e^{-{R^2 \over 2 \lambda_R^2}}(\lambda_R^2\pi)^{1/2}$,
$\lambda_r = \sqrt{{\hbar / \mu\omega_r}}$, 
and $\lambda_R = \sqrt{{\hbar / M\omega_R}}$.
By averaging ${\bf H}_\parallel$ over $\chi(y,{\bf R})$, 
we get the effective Hamiltonian 
${\bf H}_{eff} = {1 \over 2}\hbar\omega_r +\hbar\omega_R
+ {p_x^2 / 2\mu} + V_{eff}(x)$,
characterized by the effective potential 
\begin{equation}
V_{eff}(x) = {1\over 2} \mu  \omega_r^2
(x-d)^2 + V_C\left({x^2+l^2 \over 2\lambda_r^2}\right)
\label{effV}
\end{equation}
with
$V_C(u) = - (e^2 /\epsilon \sqrt{\pi} \lambda_r) 
e^{u} K_0(u)$,
$K_0$ being the zero-order Bessel function.

Since
$K_0(u)\stackrel{u\to\infty}{\sim}\sqrt{\pi/2u}\cdot e^{-u}$,  
in the
limit $x\to\infty$, we regain the expected behavior  for the Coulomb term
 \begin{equation} \label{limVC}
V_C({x^2+l^2\over 2\lambda_r^2})
\stackrel{{x^2+l^2\over 2}\gg\lambda_r^2} {\sim} - {e^2\over\epsilon
\sqrt{x^2+ l^2}}\stackrel{|x|\gg l}{\approx} - {e^2\over\epsilon
|x|}\ . 
\end{equation}

Notice  that, 
considering the typical parameters of our systems ($l\approx 20$\,\AA
\ and $\lambda_r\approx 50$\,\AA), 
according to Eq.~(\ref{limVC}) there exists a relevant range of values
for $x$ where we cannot approximate $V_C$  by its simpler asymptotic,
Coulomb-like form.

Since we
are interested in the system ground state, we can approximate 
$V_{eff}$
around its minimum: 
\begin{equation} 
V_{eff}(x) \approx V_0+{1\over 2}
\mu\tilde{\omega}^2(x-x_0)^2 \ ,
\end{equation} 
where 
$V_0\equiv V_{eff}(x_0)$ and
$\mu\tilde{\omega}^2\equiv\left.\partial^2  V_{eff}/ \partial
x^2\right|_{x_0}$. 
Within such approximation scheme, the eigenvalues and eigenfunctions of 
${\bf H}_{eff}$ can be evaluated analytically and, in particular,
the approximate ground state
eigenfunction becomes 
$ 
\Psi_{x}(x) = \left({\mu
\tilde{\omega}/\hbar \pi}\right)^{1\over 4} e^{-{1\over 2} {\mu
\tilde{\omega}\over \hbar}(x-x_0)^2} $.
In the regime we are interested in (strong
confinement and pronounced biexcitonic shift, i.e., large enough external
field),  the Coulomb attraction between electron and hole can be
regarded as a perturbation. In this regime, its main effect  is to reduce
the displacement $d$ between electron and  hole wave-function centers to
$x_0$, while the two single particle wave functions are, with a good
approximation, rigidly translated.  This
can be understood by looking at Fig.~\ref{f:Veff},
where the potential  $V_{eff}$ is
plotted for three different values of the external
field ${\bf F}$. The solid lines correspond to the full potential, 
the dashed lines
to its parabolic part, the dotted line to the Coulomb part (independent of
${\bf F}$). For small and intermediate ${\bf F}$
 the influence of the Coulomb field
on the  total potential is relevant. For intermediate fields the figure
clearly shows that the minimum of the total potential is
shifted with respect to the parabolic one. 
For strong and intermediate values of the applied 
field $F$, the effect of
the shallow Coulomb potential on the region around the minimum  of the 
total potential is mainly a {\it rigid shift} with respect to the unperturbed 
parabolic potential. For small
fields, instead,  the shape itself of the potential is definitely modified 
by the Coulomb
term. 

In the regime of interest, we can write $x_0$ as
\begin{equation}\label{r0} 
x_0=d-\Delta x, 
\end{equation}  
with 
$\Delta x\ll d$. 
By inserting (\ref{r0}) into $\partial V_{eff}/\partial x |_{x_0}=0$ 
and considering, in
the resulting equation, terms
up to first order in $\Delta x$, the following analytical expression is
obtained: 
\begin{equation}\label{delx}
{\Delta x\over d} = -{\lambda_r\over a^{ex}}{\exp(\xi)\over\sqrt{\pi}}
{\Delta K \over 1-{\lambda_r\over a^{ex}}{\exp(\xi)\over\sqrt{\pi}}
\left[{d^2\over\lambda_r^2}A(\Delta K,K_1)+\Delta K\right]}\ ,
\end{equation}
where
$\xi = (d^2+l^2)/2\lambda_r^2$, $K_1$ denotes the first-order Bessel 
function, $\Delta K=K_0(\xi)-K_1(\xi)$, $A(\Delta K,K_1)=2\Delta K+
{K_1(\xi)/ \xi}$, and $a^{ex}$
is the excitonic Bohr radius introduced in Eq.~(\ref{aex}).. 
Notice that the prefactor ${\lambda_r/ a^{ex}}$ is a measure of the 
system confinement.
In a similar way, setting $\tilde{\omega}=\omega_r+\Delta\omega$ in 
$\mu\tilde{\omega}=
\partial^2 V_{eff}/\partial x^2|_{x_0}$,
we can calculate  the effect of the Coulomb 
attraction on the potential shape:
\begin{eqnarray}
{\Delta\omega\over \omega_r}&=&
-{\lambda_r\over a^{ex}}{\exp(\xi)\over 2\sqrt{\pi}}
\left({d^2\over\lambda_r^2}A(\Delta K,K_1)+\Delta K\right.\nonumber \\
&-&\left.{\Delta x\over d}
{d^2\over\lambda_r^2}\left\{{d^2\over\lambda_r^2}\left[2A(\Delta K,K_1)-{1\over\xi}
\left(\Delta K+2{K_1(\xi)\over \xi}\right)\right]+3A(\Delta K,K_1)\right\}
\right) \ .
\end{eqnarray}
In the strong-field limit
${\lambda_r^2/ d^2}\ll 1$ 
[see Eq.~(\ref{d})], 
$ {\Delta\omega/ \omega_r}=-{\Delta x/ d}\propto -(\lambda_r/ a^{ex})
(\lambda_r^3/ d^3)$, which shows that, in this regime, Coulomb corrections
decrease very fast
with increasing field. 
The condition $\Delta x/d\stackrel{<}{\sim}20\%$ 
quantitatively defines the validity regime of the proposed approximation
scheme. The latter coincides with the intermediate- and strong-field one, 
which is the regime of interest for the QD structures investigated below.
It is also easy to show that in this regime the correction 
on the wave function due to
 $\Delta \omega/
\omega_r$ is negligible with respect to the correction given by the shift
$\Delta x/d$ 
(see also Fig.~\ref{f:num_ana}). 

As previously discussed,
the most important quantity for implementing our QIC scheme
is the biexcitonic shift. This is in our case the energy shift due to the
Coulomb interaction between two excitons sitting in neighboring dots (see 
Sect.~\ref{ss:cqd}).
Within our model, we approximate the biexcitonic ground state 
as the product of two excitonic wavefunctions sitting in 
different dots and built according to the prescription given above.
The wavefunction in the growth direction is approximated by a Gaussian of 
width $l/2$
 and the two dots are taken to have the same width $a$, 
i.e., the average of the two dots widths. This is reasonable since, for
construction, the two dots have almost the same width.
The desired biexcitonic shift $\Delta {\cal E}$ is then obtained 
averaging the corresponding two-exciton Hamiltonian over such factorized 
ground state.
Within this
approximation scheme, $\Delta {\cal E}$ becomes an easy-to-calculate sum 
of at most
2D integrals.
In the corresponding validity region 
the model provides a good estimate for $\Delta {\cal E}$: 
Figure \ref{f:num_ana} shows  a comparison between the exact results 
(diamonds),
the approximate result (solid curve) and the results obtained by neglecting 
completely Coulomb correlation in the wavefunctions (dotted line). The
dashed curve shows the approximated results obtained setting ${\Delta\omega/
\omega_r}=0$:  as anticipated before, this correction is generally 
negligible.

In order to implement our quantum computing scheme, system parameters as 
$\omega_e$, $\omega_h$ and ${\bf F}$ must satisfy  some  specific
requirements. This determines the parameter space available 
in designing our QD structures.
To this end, let us analyze the various constraints  in details.

First of all, (i) in order to have well-defined qubits, tunneling between 
dots must be suppressed;
in agreement with
state-of-the-art nanostructure technology, we have chosen for our
calculations barrier heights of $1$\,eV (electrons) and $0.58$\,eV (holes) 
and an interdot distance  $D=100$\,\AA. 
On the other side, (ii) to  implement our
QIC scheme, Coulomb interaction between consecutive dots
must be strong enough to produce a biexcitonic shift of the order of a few
meV; this can be obtained either by tailoring in a
suitable way   the distance between the two dots or by  varying
the strength of the in-plane field ${\bf F}$, since as a rough
approximation
\begin{equation}\label{DelE} 
\Delta {\cal E}\propto {d^2\over
D^3}\ ,
\end{equation} 
where $d$ is given by Eq.~(\ref{d}).
Unfortunately, (iii) a side effect of a strong electric field ${\bf F}$ is to
decrease the
oscillator strength, and, accordingly, the system response to  driving
laser  pulses; indeed, the  electric field induces a spatial separation 
between electron and hole wavefunctions,
thus decreasing their overlap 
(see Fig.~\ref{f:sqd-cd1b}).
If we now consider the confining parabolic potentials, (iv)
in order to have well-defined
quantum dots, the system must be in the strong-confinement regime 
previously introduced: the
characteristic length $\lambda_r$
associated to the parabolic potential in Eq.~(\ref{parH2}) must be smaller
than the corresponding excitonic Bohr radius $a^{ex}$. 
On the other side, (v) as shown by Eq.~(\ref{DelE}) and Eq.~(\ref{d}), 
a too strong parabolic confinement would in turn heavily decrease  
the biexcitonic shift $\Delta {\cal E}$. 
Last but not least, (vi)
in order to be able to perform 
general QIC schemes, we must be able  to energetically address specific
excitations of the system unambiguously. This means that the peaks of
interest in the optical spectra, namely ground-state excitonic and 
biexcitonic states,
must
be well-isolated from other high-energy transitions.
This determines   additional constraints  on the value of $\hbar\omega_e$
and $\hbar\omega_h$.

From the above discussion, it is clear that in
order to satisfy at the same time all the  requirements listed above 
[(i)--(vi)],
the system parameters must be fine tuned so that a quick mean to scan the 
whole parameter space becomes necessary. 
The simplified model previously described came out to be quite efficient in
performing such detailed analysis.
The available parameter space for a
reasonable field of $F=75$\,kV/cm is shown in Fig.~\ref{f:parsp}. 
Here, the typical error in the calculated values of 
$\Delta {\cal E}$ is 10-20\%. The
constraints imposed are $\Delta 
{\cal E}
\ge 3.5$\,meV, oscillator strength greater than $0.15 \mu_{bulk}$,
$\hbar\omega_e>\hbar\omega_h$,  $\hbar\omega_h-\Delta{\cal E} \ge 10$\,meV,
and $\lambda_r/a^{ex}\le 0.6$. 
Based on this analysis, we have identified the parameter set used in the 
simulated experiments of QIC processing presented in the following section.

\section{Quantum information processing}\label{s:qip}

As anticipated in the introductory part of the paper,
the advent of QIC\cite{QIC} 
as an abstract concept,
has stimulated a great deal of new thinking about how to design 
and realize quantum information processing devices.
This goal  is extremely challenging:
generally speaking, 
one should be able to perform, on a system with a well-defined 
quantum state space (the {\it computational} space), 
precise quantum-state preparation, 
coherent quantum manipulations ({\it gating}) of arbitrary length,
and state detection  as well.
It is well known that the major obstacle to implement this ideal
scheme is {\it decoherence}: the spoiling of the unitary character 
of quantum evolution
due to the uncontrollable coupling with environmental, 
i.e., non-computational, degrees of freedom.
Mostly due to the need of low decoherence  rates, 
the first proposals for experimental realizations of quantum information  
processing devices originated from specialties in atomic
physics,\cite{AP} in quantum optics,\cite{QO} and in 
nuclear and electron magnetic-resonance spectroscopy.\cite{NMR-EMR} 
On the other hand, practically relevant quantum computations require
a large number of quantum-hardware units ({\it qubits}), that
are known to be hardly  achievable in terms of such  systems. 
In contrast, in spite of the serious difficulties related to the ``fast'' 
decoherence times, a solid-state implementation of QIC
 seems to be the only way to benefit synergically from the 
recent progress in ultrafast optoelectronics\cite{UO} as well as in 
meso/nanostructure fabrication and characterization.\cite{QDs}
Among the proposed solid-state implementations one should 
mention those in superconducting-device physics\cite{SC} and in meso- and 
nanoscopic physics.\cite{QD-spin} In particular, the first 
semiconductor-based proposal, by Loss and DiVincenzo, relies on spin 
dynamics in quantum dots; it exploits the low decoherence of
spin degrees of freedom  in comparison to the one of charge excitations. 

As originally envisioned in Ref.~\onlinecite{ZR}, 
gating of charge excitations could be performed by exploiting  {\it present} 
 ultrafast laser technology,\cite{UO} that
  allows  to generate and manipulate electron-hole
quantum states on a sub-picosecond time-scale:
{\em  coherent-carrier-control}.\cite{CCC}
In this respect, decoherence times on nano/microsecond scales 
can be regarded as ``long'' ones.
Based on this idea a few implementations 
have been recently put forward;\cite{SBI}  
However, while in these 
proposals single-qubit operations are implemented  by means  of ultrafast 
optical spectroscopy,  the control of two-qubit 
operations still involves the application of external fields 
and/or microcavity-mode couplings,
whose switching times are much longer than
decoherence times in semiconductors. 
It clearly follows  that such proposals are currently out of reach in terms
of state-of-the-art optoelectronics technology.

As already pointed out in Ref.~\onlinecite{ZR}, in order to take full advantage 
from modern ultrafast laser spectroscopy 
one should be able to design fully optical gating schemes
able  to perform single- {\it and}  two-qubit operations on a 
sub-picosecond time-scale.
Following this spirit, we have recently proposed the first {\it all-optical} 
implementation with semiconductor macromolecules.\cite{PRL} 

Aim of this section is to review and discuss the semiconductor-based 
implementation in Ref.~\onlinecite{PRL}, whose quantum hardware 
consists of coupled QD structures, like those investigated in
Sect.~\ref{ss:cqd}. As described below, the crucial ingredient in our 
QIC scheme is the field-induced exciton-exciton 
coupling discussed in Sect.~\ref{s:eor}.
Indeed, the central idea in our QIC proposal is to exploit such few-exciton 
effects to design {\it  conditional operations}. 

\subsection{Quantum hardware and computational subspace}\label{ss:qhcs}

As discussed in Sect.~\ref{ss:ep}, two basic requirements are needed for QIC 
processing: (i) the tensor-product structure of the quantum hardware, and 
(ii) the $SU(2)$ character of the raising/lowering operators acting on the 
individual qubits. 
Based on the electro-optical-response analysis of Sect.~\ref{s:eor}, we can
conclude that state-of-the-art coupled QD structures can be used as 
semiconductor-based hardware for quantum information processing. 
Indeed, as shown in the previous section, these requirements
are well fulfilled by the prototypical QD molecules studied above.
Our detailed investigation has shown that a proper tailoring of the QD 
confinement potential as well as of the inter-dot distance allows to 
identify a well-precise subset of excitonic states, corresponding to 
intra-dot ground-state excitons.
Indeed, as clearly shown in Sect.~\ref{ss:cqd} (see Fig.~\ref{f:cqd-str}), 
we can associate to each QD structure a ground-state exciton, i.e., its 
low-energy optical transition corresponding to the creation/destruction of 
a Coulomb-correlated electron-hole pair in that dot.
We have shown that for these low-energy intra-dot optical transitions the 
corresponding exciton wavefunctions are localized in the various dots of
the array; this allows us to label such subset of excitonic states according to 
their host QD. 
In addition, in view of the relatively strong carrier confinement, leakage 
effects (see Sect.~\ref{ss:ep}) are expected to play a minor role.

More specifically, following the second-quantization notation, 
let us denote with 
$\vert n_\nu \rangle$ 
the absence ($n_\nu = 0$) ---no conduction-band electrons--- and the 
presence ($n_\nu = 1$) of a ground-state exciton ---a Coulomb-correlated 
electron-hole pair--- 
in dot $\nu$;
they constitute the single-qubit 
basis for the proposed QIC scheme: 
$\vert 0 \rangle_\nu$ and $\vert 1 \rangle_\nu$.
The whole  computational state-space is then spanned
by the basis set: 
\begin{equation}\label{cbs}
\vert \{n_\nu\}\rangle = 
\otimes_{\nu} \vert n_\nu \rangle, \quad (n_\nu=0,\,1) \ .
\end{equation}

The full many-body Hamiltonian ${\bf H} = {\bf H}^\circ + {\bf H}'$ in 
(\ref{td-Schr1}) restricted to the above computational space 
will be described by the following matrix elements: 
\begin{equation}\label{H_cs}
H_{\{n^{ }_\nu\}\{n'_\nu\}} = 
\left\langle \{n^{ }_\nu\} \left\vert \left({\bf H}^\circ + {\bf H}'\right) 
\right\vert \{n'_\nu\} \right\rangle 
= H^\circ_{\{n^{ }_\nu\}\{n'_\nu\}} + H'_{\{n^{ }_\nu\}\{n'_\nu\}} \ . 
\end{equation}
They are the sum of two contributions: the first one is due to the 
Coulomb-correlated carrier-system Hamiltonian; the second is due to the 
carrier-light interaction Hamiltonian in (\ref{H_prime}).
As discussed in Sect.~\ref{ss:cl}, the latter describes the 
creation/destruction of electron-hole pairs driven by ultrafast sequences 
of multicolor laser pulses. 

Let us now focus on the term $H^\circ$. As discussed in Sect.~\ref{ss:cccs},
it preserves the total number of electron-hole pairs $N$, and this is still true 
within our reduced ---i.e., computational--- subspace.
In general, the Hamilton Matrix 
$H^\circ_{\{n^{ }_\nu\}\{n'_\nu\}}$ is non-diagonal.
However, for the case of the realistic coupled QD structure analyzed in 
Sect.~\ref{ss:cqd}, non-diagonal terms are found to play a very minor role.
In this case, the latter can be neglected and the Hamiltonian matrix 
$H^\circ$ is then diagonal in our number representation $\{n_\nu\}$.
This suggests to introduce corresponding number operators acting on our 
computational subspace:
${\bf n}_\nu = 
\sum_{n_\nu = 0}^1 \vert n_\nu \rangle n_\nu \langle n_\nu \vert 
= \vert 1 \rangle_\nu \langle 1 \vert_\nu 
$.
The Hamiltonian ${\bf H}^\circ$ reduced to our computational subspace can 
then be expressed in terms of such number operators. 
In particular, for an array of coupled QDs this can be written as:
\begin{equation}\label{Htilde}
\tilde {\bf{H}}^\circ = \sum_{\nu} {\cal E}_\nu\,{\bf n}_\nu + 
{1\over 2} \sum_{\nu\nu'} \Delta{\cal E}_{\nu\nu'}\,{\bf n}_\nu\,{\bf n}_{\nu'} 
\ .
\end{equation}
Here, ${\cal E}_\nu$ denotes the energy of the ground-state exciton in dot 
$\nu$ 
while $\Delta{\cal E}_{\nu\nu'}$ is the biexcitonic shift due to the Coulomb 
interaction between dots $\nu$ and $\nu'$, introduced in Sect.~\ref{s:eor} 
[see Eq.~(\ref{DeltacalE1}) and Fig.~\ref{f:cqd-abs2}]. 

The effective Hamiltonian in (\ref{Htilde}) has exactly the same structure 
of the one proposed by Lloyd in his pioneering paper on quantum cellular 
automata,\cite{QCA} and it is the Model Hamiltonian currently used
in most of the NMR quantum-computing schemes.\cite{Cory} 
This fact is extremely important since it tells us that: 
\begin{itemize}
\item[(i) ] 
the present semiconductor-based implementation contains all 
relevant ingredients for the realization of basic QIC processing;
\item[(ii) ] 
it allows to establish a one-to-one correspondence between 
our semiconductor-based scheme and much more mature implementations, like
NMR.\cite{Cory}
\end{itemize}
\noindent
According to (\ref{Htilde}),
the single-exciton  energy ${\cal E}_\nu$ is renormalized 
by the biexcitonic shift $\Delta{\cal E}_{\nu\nu'}$, induced by the presence of
a second exciton in dot $\nu'$: 
\begin{equation}\label{Etilde}
\tilde{\cal E}_\nu = {\cal E}_\nu + 
\sum_{\nu' \ne \nu} \Delta{\cal E}_{\nu\nu'}\, n_{\nu'} \ .
\end{equation}

In order to better illustrate this  idea, 
let us focus again on the two-QD structure, 
i.e., two-qubit system, of Fig.~\ref{f:cqd-str}
and fix our attention on one of the two dots, say dot $b$.   
The effective energy gap between  $\vert 0 \rangle_b$ and $\vert 1 \rangle_b$ 
depends now on the occupation of dot $a$.
This elementary remark suggests to design properly tailored laser-pulse 
sequences to implement 
conditional logic gates between the two QD-qubits as well as
single-qubit rotations.
Indeed, by sending an ultrafast laser $\pi$-pulse with central 
energy 
$\hbar\omega_b[n_a] = {\cal E}_b + \Delta{\cal E}_{ba} n_a$, 
the transition $\vert n_b \rangle \to \vert 1-n_b \rangle$ 
($\pi$ rotation)
of the {\em target} qubit (dot $b$) is obtained if and only if the 
{\em control} qubit (dot $a$) is in the state $\vert n_a \rangle$. 
Notice that the above  scheme corresponds to the 
so-called selective population transfer in NMR;\cite{Cory} 
alternative procedures used in that field can be
adapted to the present proposal as well. 

Moreover,
by denoting with ${\cal U}_b^{n_a}$ the generic 
unitary transformation induced by
the laser $\pi$-pulse of central frequency $\omega_b[n_a]$, 
it is easy to check that the two-color pulse sequence 
${\cal U}_b^0\, {\cal U}_b^1$ 
achieves the unconditional $\pi$-rotation of qubit $b$.

\subsection{A few simulated experiments}\label{ss:afse}

In order to test the viability of the proposed quantum-computation 
strategy,
we have performed a few simulated experiments of basic quantum information 
processing. 
To this aim, we have performed a direct time-dependent solution of the 
generalized Liouville-von Neumann equation in (\ref{rho2}) restricted to 
our computational subspace, i.e., we have simulated the time evolution of 
the reduced density matrix:
\begin{equation}\label{rho1_cs}
\rho_{\{n^{ }_\nu\}\{n'_\nu\}}(t) = 
\langle \{n^{ }_\nu\} \vert \rho(t) \vert \{n'_\nu\} \rangle \ .
\end{equation}
As discussed in Sect.~\ref{ss:env}, this is governed by the total 
Hamiltonian ${\cal H}$ reduced to our computational subspace [see 
Eqs.~(\ref{H_cs}) and (\ref{Htilde})] plus a 
non-unitary term\cite{PZ} due to energy-relaxation and dephasing processes
induced by environmental degrees of freedom, like phonons, plasmons, etc. 
The latter has been treated within the standard $T_1 T_2$ model.\cite{T1-T2} 

We stress that the present density-matrix description, restricted to our 
computational subspace, does not account for {\it leakage effects}, i.e., 
it neglects processes connecting states of the computational subspace to 
other ---non-computational--- excitonic states, and viceversa.
Due to the strong-confinement character of our QD structures (see 
Sect.~\ref{s:eor}) such leakage effects are expected to play a very minor 
role. A quantitative evaluation of the leakage dynamics would require 
the inclusion in our density-matrix description of non-computational 
states.

The above simulation scheme has been applied to 
the coupled-QD
structure of Fig.~\ref{f:cqd-str} in the presence of an 
in-plane static field $F =75$\,kV/cm: 
${\cal E}_a = 1.673$\,eV,
${\cal E}_b = 1.683$\,eV,
and $\Delta{\cal E} = 4$\,meV (see inset in 
Fig.~\ref{f:cqd-abs2}).

We shall start our time-dependent analysis by simulating a basic 
conditional two-qubit operation, the so-called {\it controlled not} (CNOT) 
gate.

Our first simulated experiment is shown in Fig.~\ref{f:tds1}.
The multi-color laser-pulse train (see central panel) 
is able to perform first 
a $\pi$ rotation of the qubit $a$; Then, the second pulse is tuned 
to the frequency 
${\cal E}_b + \Delta{\cal E} $, thus performing a $\pi$ rotation 
of the qubit $b$ since
this corresponds to its renormalized transition energy 
[see Eq.~(\ref{Etilde})] 
when the neighbor qubit $a$ is in state $\vert 1 \rangle_a$.
The scenario described so far is confirmed by the time evolution of the 
exciton occupation numbers $n_a$ and $n_b$ (upper panel)
as well as of the diagonal elements of the density matrix
in our four-dimensional computational basis (lower panel).

More specifically, 
at the beginning the system is in the state $\vert 0,0 \rangle \equiv
\vert 0 \rangle_a \otimes \vert 0 \rangle_b$. Due to the first pulse at 
$t = 0$ the computational state moves to the state $\vert 1,0 \rangle 
\equiv \vert 1 \rangle_a \otimes \vert 0 \rangle_b$. 
Finally, at time $t = 1$\,ps the second pulse brings the system into the state
$\vert 1,1 \rangle \equiv \vert 1 \rangle_a \otimes \vert 1 \rangle_b$.   

This realizes the first part of the well-known CNOT gate: 
the target qubit $b$ is rotated if 
the control qubit $a$ is in state $\vert 1 \rangle_a$. 
To complete it, one has to show that 
the state of the target qubit $b$ remains unchanged if the control qubit 
$a$ is 
in state $\vert 0 \rangle_a$.
This has been checked by a second simulated experiment (not reported here) 
where the first pulse, being now off-resonant (with 
respect to dot $a$), does not change the computational state of the system. 
As a consequence, the second pulse is no more into resonance with the 
excitonic-transition energy of dot $b$, since the latter is no more 
renormalized by the excitonic occupation of dot $a$.
Therefore, the initial state of the system is
$\vert 0,0 \rangle$ and the final one is again $\vert 0,0 \rangle$.

The simulated experiments discussed so far clearly 
show the potential realization of the CNOT gate, thus confirming 
the validity of the proposed semiconductor-based QIC strategy.
However, the analysis presented so far deals with factorized states, 
i.e., we have simulated the CNOT gate acting on basis states 
$\vert \{n_\nu\} \rangle$ only.
It is well known\cite{QIC} that the key ingredient in any 
quantum-computation protocol is entanglement. Generally speaking, this 
corresponds to a non-trivial linear combination of our basis states. 

We shall now show that the CNOT gate previously discussed is able to transform 
a factorized state into a maximally entangled one.
Figure \ref{f:tds2} shows a simulated two-qubit operation 
driven again by a two-color laser-pulse sequence (see central panel).
Initially,
the system is in the state $\vert 0,0 \rangle$. 
The first laser pulse (at $t = 0$) is tailored 
in such a way to induce now a $\pi\over 2$ rotation of the qubit 
$a$: 
$\vert 0,0 \rangle \to (\vert 0,0 \rangle + \vert 1,0 \rangle)/\sqrt{2}$. 
At time $t = 1$\,ps a second pulse induces a conditional $\pi$-rotation of 
the qubit $b$:
$\vert 0,0 \rangle + \vert 1,0 \rangle \to 
\vert 0,0 \rangle + \vert 1,1 \rangle$.
This last operation plays a central role in any QIC
processing, since it transforms a {\it factorized} state 
($(\vert 0 \rangle_a + \vert 1 \rangle_a) \otimes \vert 0 \rangle_b$) 
into an  {\it entangled} state 
($\vert 0 \rangle_a \otimes \vert 0 \rangle_b + 
\vert 1 \rangle_a \otimes \vert 1 \rangle_b$).

As we can see, during the pulse energy-nonconserving (or off-resonant) 
transitions\cite{SST} take place; however, at the end of the pulse 
such effects vanish and the desired quantum state is reached.

The experiments simulated above 
(see Figs.~\ref{f:tds1} and \ref{f:tds2})
clearly show that the energy scale of the 
biexcitonic splitting $\Delta{\cal E}$ in our QD molecule (see 
Fig.~\ref{f:cqd-abs2}) 
is compatible with the sub-picosecond operation time-scale of 
modern ultrafast laser technology.\cite{UO}

\section{Summary and conclusions}\label{s:conc}

We have presented a detailed analysis of the electro-optical response of 
single as well as coupled QD structures. More specifically, we have 
investigated the effect of a static electric field on the many-exciton 
optical response of quasi 0D semiconductor nanostructures.
Our analysis has shown that a proper tailoring of the single-particle 
confinement potential as well as of the inter-dot distance and applied 
field allows to induce and control intra- as well as interdot 
exciton-exciton coupling; this, in turn, may give rise to significant 
energy shifts of the optical transitions.

This field-induced dipole-dipole coupling constitutes the key ingredient of
the proposed all optical implementation of QIC with 
a semiconductor-based quantum hardware. 
Our analysis has
shown that energy-selected optical transitions in realistic state-of-the-art 
QD structures are good candidates for quantum-information encoding and 
manipulation. The sub-picosecond time-scale of ultrafast laser spectroscopy
allows for a relatively large number of elementary operations within the 
exciton decoherence time.

At this point a few comments are in order.
First, we stress a very important feature of the proposed semiconductor-based 
implementation:
as for NMR quantum computing, 
two-body interactions 
are always switched on 
(this  should be compared to the schemes in which 
two-qubit gates are realized by turning on and off the coupling between 
subsystems, e.g., by means of slowly-varying fields and cavity-mode 
couplings);
conditional as well as unconditional dynamics
is realized by means of sequences of ultrafast  single-qubit operations 
whose length 
does not scale as a function of the total number of QDs in the 
array.\cite{NN}

Let us now come to the {\it state measurement}. 
In view of the few-exciton character of the proposed quantum hardware,
the conventional measurement of the carrier subsystem 
by spectrally-resolved luminescence needs to be replaced by more 
sensitive detection schemes. 
To this end, a viable strategy could be to apply to our semiconductor-based
structure the well-known recycling techniques commonly used in 
quantum-optics experiments.\cite{recycling}
Generally speaking, the idea is to properly combine quantum- and 
dielectric-confinement effects in order to obtain well-defined energy levels, 
on which  design energy-selective photon-amplification schemes. 
An alternative approach would be to adopt a {\it storage-qubit scheme}, as 
recently proposed in Ref.~\onlinecite{storage}.

The nanoscale range of the inter-dot coupling we employed for 
enabling conditional dynamics 
does not allow for space-selective
optical addressing of individual qubits. 
For this reason, at least for our basic QD molecule ($a + b$), we resorted to 
an energy-selective addressing scheme. 
However, extending such strategy to the whole QD array would imply 
different values of the excitonic transition in each QD, 
i.e., ${\cal E}_\nu\neq {\cal E}_{\nu'}$.  
This, besides obvious technological difficulties, would constitute a conceptual
limitation  of  scalability  towards  massive Quantum Computations.
The problem can be avoided following a completely different strategy 
originally proposed by Lloyd\cite{QCA} and recently improved in 
Ref.~\onlinecite{Benjamin}: 
by properly designed sequences of multicolor global pulses within a 
cellular-automaton scheme, local addressing is replaced by 
information-encoding transfer along our QD array.

Finally, a present limitation of the proposed quantum hardware are the 
non-uniform structural and geometrical properties of the QDs in the array, 
which may give rise to energy broadenings larger than the biexcitonic 
shift. 
However, recent progress in QD fabrication
---including the realization of QD structures in microcavities--- 
will allow, we believe, to overcome this purely technological 
limitation.

\section*{Acknowledgments}

We are grateful to David DiVincenzo, Rita Iotti, Neil Johnson, Seth Lloyd, 
Ehoud Pazy, and Peter Zoller for stimulating and fruitful discussions.

This work has been supported in part by the European Commission through the
Research Project {\it SQID} within the {\it Future and Emerging 
Technologies (FET)} programme.

\begin{appendix}

\section{Evaluation of single-particle properties}\label{app:spp}

In this section we shall describe the numerical approach used for the 
evaluation of the single-particle properties ---3D wavefunctions and 
energy levels--- for single as well as coupled QD structures.
Within the standard envelope-function picture,\cite{Bastard} 
the non-interacting carriers in our quasi-0D
structure in the presence of a static electric field are described by the 
Schr\"odinger equation (\ref{Schr1}) with the 
confinement potential in (\ref{F}):
\begin{equation}\label{A:Schr1}
\left[
-{\hbar^2\nabla^2_{\bf r}\over 2 m_{e/h}} 
+ V_c^{e/h}({\bf r})
\pm e {\bf F \cdot r}
\right] \psi_{i/j}({\bf r}) = \epsilon_{i/j} \psi_{i/j}({\bf r}) \ .
\end{equation}

As for the case of semiconductor quantum wires,\cite{QWR} 
a quantitative analysis
of the whole single-particle spectrum $\epsilon_{i/j}$ 
requires a direct numerical solution of 
the above Schr\"odinger equation. 
This can be performed using a 
fully 3D plane-wave expansion described in Ref.~\onlinecite{Troiani}, which 
is a straightforward generalization to QD structures of the 2D 
plane-wave expansion proposed in Ref.~\onlinecite{QWR}.

As anticipated in Sect.~\ref{s:eor}, 
when ---as in this paper--- we are interested in the low-energy range only,
for most of the QD structures realized so far the carrier confinement 
can be described as the sum of two potential profiles acting along different 
directions [see Eq.~(\ref{V_c})], 
which allows us to factorize the original 3D problem in (\ref{A:Schr1}) into a
perpendicular ($\perp$) direction and a parallel ($\parallel$) plane [see 
Eqs.~(\ref{perpar1}) and (\ref{perpar2})].
Moreover, as far as the low-energy 
region is concerned, the in-plane or parallel confinement is well described 
by a 2D parabolic potential. 
In this case the Schr\"odinger equation within the 2D parallel subspace
can be solved analytically (see below)
and thus our problem reduces to a numerical 
solution of the Schr\"odinger equation along the perpendicular direction:
\begin{equation}\label{A:Schr2}
{\cal H} \psi^\perp_{i_\perp/j_\perp}(r_\perp) = 
\left[
-{\hbar^2\nabla^2_{r_\perp} \over 2 m_{e/h}} 
+ V_\perp^{e/h}(r_\perp)
\pm e F_\perp r_\perp
\right] \psi^\perp_{i_\perp/j_\perp}(r_\perp) = 
\epsilon^\perp_{i_\perp/j_\perp} \psi^\perp_{i_\perp/j_\perp}(r_\perp) \ .
\end{equation}
This has been solved using the plane-wave-expansion technique previously 
discussed.\cite{QWR,Troiani}
Within such approach, the unknown envelope function is written as a linear 
combination of plane waves, i.e.,
\begin{equation}\label{A:pw1}
\psi^\perp_{i_\perp/j_\perp}(r_\perp) = {1 \over \sqrt{L}} \sum_G 
b_G e^{i G r_\perp} \ ,
\end{equation}
where $G = n {2\pi \over L}$ are reciprocal lattice vectors associated to the 
periodicity box $L$.
By substituting the above plane-wave expansion into Eq.~(\ref{A:Schr2}), 
the latter is transformed into the following eigenvalue problem:
\begin{equation}\label{A:pw2}
\sum_G \left({\cal H}_{GG'} -\epsilon^\perp \delta_{GG'} 
\right) b_{G'} = 0 \ ,
\end{equation}
where ${\cal H}_{GG'}$ are the matrix elements of the single-particle
Hamiltonian in (\ref{A:Schr2}) within our plane-wave basis.
A direct diagonalization of ${\cal H}_{GG'}$ will then provide the desired
perpendicular energy levels $\epsilon^\perp$ as well as the wavefunction 
coefficients 
$b_G$.

Let us now come back to the in-plane or parallel-subspace problem, which we
treat within the 2D parabolic-confinement model previously mentioned, i.e.,
\begin{equation}\label{A:pcm1}
V^{e/h}_\parallel({\bf r}_\parallel) = {1 \over 2} k_{e/h} r^2_\parallel \ 
.
\end{equation}
The corresponding Schr\"odinger equation is of the form:
\begin{equation}\label{A:Schr3}
\left[
-{\hbar^2\nabla^2_{{\bf r}_\parallel} \over 2 m_{e/h}} 
+ {1 \over 2} k_{e/h} r^2_\parallel
\pm e {\bf F}_\parallel \cdot {\bf r}_\parallel
\right] \psi^\parallel_{i_\parallel/j_\parallel}({\bf r}_\parallel) = 
\epsilon^\parallel_{i_\parallel/j_\parallel} 
\psi^\parallel_{i_\parallel/j_\parallel}({\bf r}_\parallel) \ .
\end{equation}
It is well known that the presence of a static uniform electric field
${\bf F}_\parallel$ does not change the parabolic nature of our confinement
potential. Indeed, Eq.~(\ref{A:Schr3}) can be rewritten as:
\begin{equation}\label{A:Schr4}
\left[
-{\hbar^2\nabla^2_{{\bf r}_\parallel} \over 2 m_{e/h}} 
+ {1 \over 2} k_{e/h} \vert {\bf r}_\parallel - {\bf d}^{e/h}_\parallel\vert^2
\right] \psi^\parallel_{i_\parallel/j_\parallel}({\bf r}_\parallel) = 
\left(
\epsilon^\parallel_{i_\parallel/j_\parallel} - \Delta\epsilon_{e/h}
\right)
\psi^\parallel_{i_\parallel/j_\parallel}({\bf r}_\parallel) \ .
\end{equation}
The presence of the applied field results in a shift 
\begin{equation}\label{A:d_eh}
{\bf d}^{e/h}_\parallel = \mp {e{\bf F}_\parallel \over k_{e/h}}
\end{equation}
of the parabolic-potential minimum as well as in a rigid energy shift
\begin{equation}\label{A:Deltaepsilon}
\Delta\epsilon_{e/h} = -{1 \over 2} k_{e/h} {{\bf d}^{e/h}_\parallel}^2\ ,
\end{equation}
often referred to as Stark shift. 
We stress that in the presence of the electric field ${\bf F}_\parallel$ we
have different symmetry centers for electrons and holes; this, in turn, 
introduces significant modifications in the selection rules governing 
interband optical transitions (see below).

As anticipated, the Schr\"odinger equation (\ref{A:Schr4}) can be solved 
analytically. Due to the central symmetry of the problem (with respect to 
the parabolic-potential minimum ${\bf d}^{e/h}_\parallel$),
it is convenient to adopt a 2D polar-coordinate set.
By denoting with 
${\sl r} = \vert {\bf r}_\parallel - {\bf d}^{e/h}_\parallel \vert$ the 
radial coordinate and with $\varphi$ the azimuthal coordinate measured with 
respect to the field direction,
we have:
\begin{equation}\label{A:psipar}
\psi^\parallel_{nm}\left({\sl r},\varphi\right) =
\alpha^{-(\left|m\right|+1)}\sqrt{\frac{n!}{\pi
\left(n+\left|m\right|\right)!}} e^{-im\varphi}
{\sl r}^{\left|m\right|}
e^{-\frac{ {\sl r}^2}{2\alpha^2}}
{\cal L}_n^{\left|m\right|}\!\left( \frac{{\sl r}^2}{\alpha^2}\right) \ ,
\end{equation}
where
\begin{equation}\label{A:alpha}
\alpha = 
\left(\frac{m_{e/h}\omega_{e/h}}{\hbar}\right)^{-{1 \over 2}}
\end{equation}
is the spatial extension of the harmonic-oscillator ground state 
---$\omega_{e/h} = \sqrt{k_{e/h} \over m_{e/h}}$ being its oscillation 
frequency--- while
${\cal L}_n^{\left|m\right|}\left(x\right)$ denotes the generalized Laguerre 
polynomial in the dimensionless variable
$x = {\sl r}^2/\alpha^2$. 

In view of the central symmetry of the problem, our quantum numbers are 
those of the angular momentum in two dimensions, i.e.,
a radial number $n$ ($n = 0, 1, 2, \dots$) plus an orbital number $m$
($m = -n, -n+2, \dots, n-2, n$).

The corresponding in-plane single-particle energy spectrum is given by:
\begin{equation}\label{A:epsilon_par}
\epsilon^\parallel_{nm} = 
\hbar\omega_{e/h} \left(2n - \left|m\right| + 1 \right) =
\hbar\omega_{e/h}\left(n_\epsilon + 1\right) \ ,
\end{equation}
where 
$n_\epsilon = 2n - |m|$  
denotes the energy quantum number with degeneracy 
$(n_\epsilon + 1)$.

The 3D single-particle energy spectrum is then given by the sum of 
equally-spaced energy-level sequences, i.e.,
\begin{equation}\label{A:epsilon}
\epsilon_l = \epsilon^\perp_{l_\perp} + \epsilon^\parallel_{nm} = 
\epsilon^\perp_{l_\perp} + (n_\epsilon+1) \hbar\omega_{e/h} \ :
\end{equation}
for each energy level $\epsilon^\perp$ ---obtained by solving the eigenvalue
problem in (\ref{A:pw2})--- we deal with an harmonic-oscillator spectrum with 
energy separation $\hbar\omega_{e/h}$.

Given the single-particle state factorization in (\ref{perpar1}), the 
corresponding dipole matrix elements in Eq.~(\ref{mu}) can be factorized as
well:
\begin{equation}\label{A:mu}
\mu_{ij} = \mu_{bulk} 
{\cal I}^\perp_{i_\perp j_\perp} 
{\cal I}^\parallel_{i_\parallel j_\parallel} \ ,
\end{equation}
with
\begin{equation}\label{A:muper1}
{\cal I}^\perp_{i_\perp j_\perp} = \int \psi^\perp_{i_\perp}(r_\perp) 
\psi^\perp_{j_\perp}(r_\perp) dr_\perp
\end{equation}
and
\begin{equation}\label{A:mupar1}
{\cal I}^\parallel_{i_\parallel j_\parallel} = \int 
\psi^\parallel_{i_\parallel}({\bf r}_\parallel) 
\psi^\parallel_{j_\parallel}({\bf r}_\parallel) d{\bf r}_\parallel \ .
\end{equation}
By inserting the plane-wave expansion (\ref{A:pw1}) into 
Eq.~(\ref{A:muper1}), we get:
\begin{equation}\label{A:muper2}
{\cal I}^\perp_{i_\perp j_\perp} = \sum_G b^{i_\perp}_G b^{j_\perp}_G \ .
\end{equation}

Let us finally focus on the in-plane integral in (\ref{A:mupar1}). This can 
be rewritten in terms of the polar-coordinate set ${\sl r}, \varphi$ 
introduced in (\ref{A:psipar}):
\begin{equation}\label{A:mupar2}
{\cal I}^\parallel_{nm,n'm'} = \int 
\psi^\parallel_{nm}({\sl r}_e, \varphi_e) 
\psi^\parallel_{n'm'}({\sl r}_h, \varphi_h) d{\bf r}_\parallel \ .
\end{equation}
Here, 
${\sl r}_{e/h} = \vert {\bf r}_\parallel - {\bf d}^{e/h}_\parallel \vert$ 
and $\varphi_{e/h}$ are the corresponding azimuthal angles.
In general, the two polar-coordinate sets for electrons and holes do not 
coincide. Indeed, in the presence of a static field $F_\parallel$ we have 
different symmetry centers [see Eq.~(\ref{A:d_eh})].
In contrast, for $F_\parallel = 0$ the two coordinate sets coincide 
(${\bf r}_\parallel \equiv {\sl r}, \varphi = {\sl r}_e, \varphi_e = {\sl 
r}_h, \varphi_h$) and the above equation reduces to:
\begin{equation}
{\cal I}^\parallel_{nm,n'm'} = \int 
\psi^\parallel_{nm}({\sl r}, \varphi) 
\psi^\parallel_{n'm'}({\sl r}, \varphi) \, {\sl r} d{\sl r} d\varphi \ .
\end{equation}
In this case ---for which the symmetry 
centers for electrons and holes coincide---
we deal with a number of well-known selection rules. In particular we have:
\begin{equation}\label{A:amc}
{\cal I}^\parallel_{nm,n'm'} \propto \delta_{m+m'} \ .
\end{equation}
This tells us that in the electron-hole generation process the
total angular momentum is conserved. 
Moreover, for the special case of equally extended electron and hole 
wavefunctions, i.e., $\alpha_e = \alpha_h$ (see discussion in 
Sect.~\ref{ss:sqd}), we have
\begin{equation}\label{A:amc-bis}
{\cal I}^\parallel_{nm,n'm'} \propto 
\delta_{m+m'} \delta_{n,n'} = 
\delta_{m+m'} \delta_{n_\epsilon,n'_\epsilon}  \ ,
\end{equation}
i.e., the energy quantum number $n_\epsilon$ is conserved as well.

In contrast, 
in the presence of the static field the above selection rules are 
violated, due to the fact that the Harmonic-oscillator 
wavefunctions in (\ref{A:psipar}) are no longer eigenstates of the 
total angular momentum.
As we shall see, the same considerations apply to the
case of the two-body Coulomb matrix 
elements discussed in the following section.

\section{Evaluation of two-body Coulomb matrix elements} \label{app:cme}

In this section we shall describe the numerical approach used for the 
evaluation of the two-body Coulomb matrix elements.
Starting again from the single-particle state factorization in 
(\ref{perpar1}), the 
Coulomb matrix elements in Eq.~(\ref{V_cc}) can be rewritten as:
\begin{equation}\label{B:cme1}
V_{l_1' l_2' l_2 l_1} = 
\int d{\bf r}_\parallel \int d{\bf r}'_\parallel 
\psi^{\parallel *}_{n_1'm_1'}({\bf r}_\parallel)
\psi^{\parallel *}_{n_2'm_2'}({\bf r}'_\parallel)
V^\parallel_{{\rm l}_1'{\rm l}_2' {\rm l}_2{\rm l}_1}
({\bf r}_\parallel-{\bf r}'_\parallel)
\psi^\parallel_{n_2m_2}({\bf r}'_\parallel)
\psi^\parallel_{n_1m_1}({\bf r}_\parallel) \ ,
\end{equation}
where 
\begin{equation}\label{B:cme2}
V^\parallel_{{\rm l}_1'{\rm l}_2' {\rm l}_2{\rm l}_1}
({\bf r}_\parallel - {\bf r}'_\parallel) = 
\int d{\bf r}_\perp \int d{\bf r}'_\perp 
\psi^{\perp *}_{{\rm l}_1'}(r_\perp)
\psi^{\perp *}_{{\rm l}_2'}(r'_\perp)
V({\bf r-r'})
\psi^\perp_{{\rm l}_2}(r'_\perp)
\psi^\perp_{{\rm l}_1}(r_\perp) 
\end{equation}
can be regarded as an in-plane effective potential obtained by integrating the 
original 3D Coulomb potential $V$ ---multiplied by the 
corresponding wavefunctions $\psi^\perp$---
over the perpendicular direction.

Let us consider the explicit form of the 3D Coulomb potential in 
(\ref{B:cme2}) written in terms of its Fourier transform along the 
perpendicular direction:
\begin{equation}\label{B:cme3}
V({\bf r - r'}) = 
{e^2 \over \varepsilon_0 \left\vert {\bf r - r'} \right\vert} =
{e^2 \over \varepsilon_0 \pi}\,\int dq 
K_0\left(q\left\vert {\bf r}_\parallel - {\bf r}'_\parallel \right\vert \right)
e^{i q \left(r_\perp-r'_\perp\right)}
\ .
\end{equation}
Here, $\varepsilon_0$ is the static dielectric constant,\cite{screening}
$q$ denotes the Fourier-transform parameter, while
\begin{equation}\label{B:K-0}
K_0(x) =
\int_0^\infty {\cos y\,dy \over \sqrt{x^2+y^2}}
\end{equation}
is the zero-order modified Bessel function.

By inserting the Fourier expansion (\ref{B:cme3}) 
into Eq.~(\ref{B:cme2}), we realize a factorization of the two space 
coordinates $r_\perp$ and $r'_\perp$. 
Indeed, by introducing the form 
factors
\begin{equation}\label{B:cme4}
{\cal F}_{{\rm ll'}}(q) =
\int dr_\perp 
\psi^{\perp *}_{{\rm l}}(r_\perp)
e^{iqr_\perp}
\psi^\perp_{{\rm l}'}(r_\perp) \ ,
\end{equation}
the Coulomb matrix elements in 
(\ref{B:cme2}) 
can be simply written as:
\begin{equation}\label{B:cme5}
V^\parallel_{{\rm l}_1'{\rm l}_2' {\rm l}_2{\rm l}_1}
({\bf r}_\parallel - {\bf r}'_\parallel) = 
{e^2 \over \varepsilon_0 \pi}\,\int 
K_0\left(q\left\vert {\bf r}_\parallel - {\bf r}'_\parallel\right\vert\right)
{\cal F}^{ }_{{\rm l}'_1 {\rm l}_1}(q) 
{\cal F}^{*}_{{\rm l}'_2 {\rm l}_2}(q) dq\ .
\end{equation}
Therefore, the evaluation of the effective Coulomb potential in 
(\ref{B:cme2}) 
reduces to the evaluation of the form factors ${\cal F}$ in Eq.~(\ref{B:cme4}).
To this end, by replacing the wavefunctions $\psi^\perp$ with their 
plane-wave expansion in (\ref{A:pw1}) we get:
\begin{equation}\label{B:cme6}
{\cal F}_{{\rm ll}'}(q) = 
\sum_{GG'} b^{{\rm l} *}_G b^{{\rm l}'}_{G'} O(G'-G+q) \ ,
\end{equation}
where
\begin{equation}
O(k) = {1 \over L} \int e^{ikr_\perp} dr_\perp
\end{equation}
are plane-wave overlap integrals over the periodicity region $L$, whose 
explicit form can be evaluated analytically. 

Therefore, for any shape of the perpendicular confinement potential, 
starting from the 
numerically computed eigenvectors $b_G$ 
[see Eqs.~(\ref{A:pw1}) and (\ref{A:pw2})],
we are able to obtain the various
form factors ${\cal F}$ which, in turn, allow us to numerically evaluate the 
effective in-plane Coulomb potential in (\ref{B:cme2}).
Once the latter is known over a suitable space grid, 
the original six-dimensional integral in (\ref{V_cc}) is 
then reduced to the evaluation of the four-dimensional integral in 
(\ref{B:cme1}).
This requires some care, since the effective potential $V^\parallel$ is 
singular for 
$\left\vert{\bf r}_\parallel -{\bf r}'_\parallel\right\vert = 0$.
In order to eliminate such singularity, it is convenient to replace the 
integration coordinate ${\bf r}'_\parallel$ with the relative coordinate 
$\overline{\bf r}_\parallel = {\bf r}_\parallel -{\bf r}'_\parallel$.
Indeed, if we move to 2D polar-coordinate sets for the new integration 
variables ${\bf r}_\parallel$ and $\overline{\bf r}_\parallel$, the 
presence of the Jacobian function corresponding to the relative coordinate 
$\overline{\bf r}_\parallel$ cancels the potential singularity.

We stress that ---as for the case of optical matrix elements previously 
discussed (see App.~\ref{app:spp})---
 in the absence of applied static fields the symmetry centers for electrons
and holes coincide [see Eq.~(\ref{A:d_eh})] and, due to global rotation 
symmetry, we get:
\begin{equation}\label{B:cme7}
V_{l_1' l_2' l_2 l_1} \propto \delta_{m_1+m_2,m_1'+m_2'}  
\end{equation}
and the numerical integration in Eq.~(\ref{B:cme1}) reduces to three 
variables only: one angular and two radial coordinates.
As for the optical matrix elements previously discussed, the selection 
rule in (\ref{B:cme7}) describes the conservation of the total angular 
momentum in the Coulomb interaction process: $m_1+m_2 = m_1'+m_2'$.

In contrast, in the presence of an applied static field the selection rule 
(\ref{B:cme7}) is relaxed and we need to numerically solve the 
four-dimensional integral in Eq.~(\ref{B:cme1}). 

\section{Evaluation of many-exciton states and optical matrix elements}
\label{app:emb}

In this section we shall apply the exact-diagonalization approach 
introduced in Sect.~\ref{ss:cccs} to the excitonic ($N = 1$) and 
biexcitonic ($N = 2$) case.
Generally speaking, the method consists in a numerical diagonalization of the
interacting-carrier Hamiltonian ${\bf H}^\circ$ written in the 
single-particle basis $\{\vert l_N \rangle\}$ [see Eqs.~(\ref{Schr3}), 
(\ref{ket_lambdaN}), and (\ref{ep})].

For the evaluation of excitonic states, i.e., states corresponding to a single 
Coulomb-correlated electron-hole pair, the proper basis set is given by the
single-particle states in (\ref{ket_lN}) with $N = 1$, i.e.,
\begin{equation}\label{C:ket_l1}
\vert l_1 \rangle \equiv 
\vert i_1 j_1 \rangle =
c^\dagger_{i_1} d^\dagger_{j_1} \vert 0 \rangle \ .
\end{equation}
The corresponding Hamiltonian matrix is given by:
\begin{equation}\label{C:H_0-l1}
H^\circ_{i_1j_1,i_1'j_1'} = 
H^c_{i_1j_1,i_1'j_1'} + 
H^{cc}_{i_1j_1,i_1'j_1'} 
\end{equation}
with
\begin{equation}\label{C:H_c-l1}
H^c_{i_1j_1,i_1'j_1'} = \langle i_1j_1 \vert {\bf H}^c \vert i_1'j_1' 
\rangle
\end{equation}
and
\begin{equation}\label{C:H_cc-l1}
H^{cc}_{i_1j_1,i_1'j_1'} = 
\langle i_1j_1 \vert {\bf H}^{cc} \vert i_1'j_1' \rangle \ .
\end{equation}
Combining Eq.~(\ref{C:ket_l1}) with the explicit form of the 
non-interacting Hamiltonian ${\bf H}^c$ in (\ref{H_c}) and making use of the
Fermionic commutation relations we get:
\begin{equation}
H^c_{i_1j_1,i_1'j_1'} = \left(\epsilon_{i_1} + \epsilon_{j_1}\right) 
\delta_{i_1j_1,i_1'j_1'} \ .
\end{equation}
In a similar way, combining Eq.~(\ref{C:ket_l1}) with the explicit form of the 
carrier-carrier Hamiltonian ${\bf H}^{cc}$ in (\ref{H_cc}), after a 
straightforward calculation we obtain:
$
H^{cc}_{i_1j_1,i_1'j_1'} = - V_{i_1j_1 j'_1 i'_1} 
$.

Let us now come to the evaluation of biexcitonic states, i.e., states 
corresponding to two Coulomb-correlated electron-hole pairs. 
In this case the proper basis set is given by the
single-particle states in (\ref{ket_lN}) with $N = 2$, i.e.,
\begin{equation}\label{C:ket_l2}
\vert l_2 \rangle \equiv 
\vert i_1j_1i_2j_2 \rangle =
c^\dagger_{i_1} d^\dagger_{j_1} 
c^\dagger_{i_2} d^\dagger_{j_2} 
\vert 0 \rangle \ .
\end{equation}
The corresponding Hamiltonian matrix is given by:
\begin{equation}\label{C:H_0-l2}
H^\circ_{i_1j_1i_2j_2,i_1'j_1'i_2'j_2'} = 
H^c_{i_1j_1i_2j_2,i_1'j_1'i_2'j_2'} + 
H^{cc}_{i_1j_1i_2j_2,i_1'j_1'i_2'j_2'}
\end{equation}
with
\begin{equation}\label{C:H_c-l2}
H^c_{i_1j_1i_2j_2,i_1'j_1'i_2'j_2'} = 
\langle i_1j_1 i_2j_2 \vert {\bf H}^c \vert i_1'j_1' i_2'j_2' \rangle
\end{equation}
and
\begin{equation}\label{C:H_cc-l2}
H^{cc}_{i_1j_1i_2j_2,i_1'j_1'i_2'j_2'} = 
\langle i_1j_1 i_2j_2 \vert {\bf H}^{cc} \vert i_1'j_1' i_2'j_2' \rangle \ .
\end{equation}
Again, combining Eq.~(\ref{C:ket_l2}) with the explicit form of the 
non-interacting Hamiltonian ${\bf H}^c$ in (\ref{H_c}) and making use of the
Fermionic commutation relations, in this case we get:
\begin{eqnarray}
H^c_{i_1j_1i_2j_2,i_1'j_1'i_2'j_2'} &=& \left(\epsilon_{i_1} + \epsilon_{i_2} +
\epsilon_{j_1} + \epsilon_{j_2} \right) \delta_{i_1j_1,i_1'j_1'}
\delta_{i_2j_2,i_2'j_2'} \nonumber \\
& - &   \left(\epsilon_{i_1} + \epsilon_{i_2} + 
\epsilon_{j_1} + \epsilon_{j_2} \right) \delta_{i_1j_1,i_1'j_2'}
\delta_{i_1j_2,i_2'j_1'}  \nonumber \\
& - & \left(\epsilon_{i_1} + \epsilon_{i_2} +
\epsilon_{j_1} + \epsilon_{j_2} \right) \delta_{i_1j_1,i_2'j_1'}
\delta_{i_2j_2,i_1'j_2' } \nonumber \\
 & +& 
\left(\epsilon_{i_1} + \epsilon_{i_2} +
\epsilon_{j_1} + \epsilon_{j_2} \right) \delta_{i_1j_1,i_2'j_2'}
\delta_{i_2j_2,i_1'j_1'}
\ .
\end{eqnarray}
In a similar way, combining Eq.~(\ref{C:ket_l2}) with the explicit form of the 
carrier-carrier Hamiltonian ${\bf H}^{cc}$ in (\ref{H_cc}), after a 
straightforward calculation we obtain:
\begin{eqnarray}
H^{cc}_{i_1j_1i_2j_2,i_1'j_1'i_2'j_2'} &=&
\phantom{+} \frac{1}{2} \left( V_{i_1 i_2 i_1' i_2'} -  V_{i_2 i_1 i_1' i_2'}
-   V_{i_1 i_2 i_2' i_1'} +  V_{i_2 i_1 i_2' i_1'}\right)
\left( \delta_{j_1j_2,j_1'j_2'} - \delta_{j_1j_2,j_2'j_1'} \right)
  \nonumber \\   
& & + \frac{1}{2} \left( V_{j_1 j_2 j_1' j_2'} -  V_{j_2 j_1 j_1' j_2'}
-   V_{j_1 j_2 j_2' j_1'} +  V_{j_2 j_1 j_2' j_1'}\right)
\left( \delta_{i_1i_2,i_1'i_2'} - \delta_{i_1i_2,i_2'i_1'} \right)
  \nonumber \\   
& & 
- \phantom{\frac{1}{2} } V_{i_1j_1i_1'j_1'} \delta_{i_2j_2,i_2'j_2'}  
+  V_{i_1j_1i_1'j_2'} \delta_{i_2j_2,i_2'j_1'}
+  V_{i_1j_2i_1'j_1'} \delta_{i_2j_1,i_2'j_2'}
-  V_{i_1j_2i_1'j_2'} \delta_{i_2j_2,i_1'j_1'} \nonumber \\
& & 
+ \phantom{\frac{1}{2}} V_{i_1j_1i_2'j_1'} \delta_{i_2j_2,i_1'j_2'} 
-  V_{i_1j_1i_2'j_2'} \delta_{i_2j_2,i_1'j_1'}
-  V_{i_1j_2i_2'j_1'} \delta_{i_2j_1,i_1'j_2'}
+  V_{i_1j_2i_2'j_2'} \delta_{i_2j_1,i_1'j_1'} \nonumber \\
& & 
+  \phantom{\frac{1}{2}} V_{i_2j_1i_1'j_1'} \delta_{i_1j_2,i_2'j_2'}
-  V_{i_2j_1i_1'j_2'} \delta_{i_1j_2,i_2'j_1'}
-  V_{i_2j_2i_1'j_1'} \delta_{i_1j_1,i_2'j_2'}
+  V_{i_2j_2i_1'j_2'} \delta_{i_1j_1,i_2'j_1'} \nonumber \\
& & 
-  \phantom{\frac{1}{2}} V_{i_2j_1i_2'j_1'} \delta_{i_1j_2,i_1'j_2'}
+  V_{i_2j_1i_2'j_2'} \delta_{i_1j_2,i_1'j_1'}
+  V_{i_2j_2i_2'j_1'} \delta_{i_1j_1,i_1'j_2'}             
-  V_{i_2j_2i_2'j_2'} \delta_{i_1j_1,i_1'j_1'}\ . 
\end{eqnarray}

Let us finally discuss the explicit form of the carrier-light matrix elements 
(\ref{H_lambda_lambdaprime}) entering the many-exciton absorption probability 
in Eq.~(\ref{FGR1}).

For the excitonic absorption [see Eq.~(\ref{FGR2})]
the corresponding matrix elements are defined in Eq.~(\ref{H_lambda1_0}).
Combining the explicit form of the carrier-light interaction Hamiltonian 
(\ref{H_prime}) with that of the generic excitonic state in 
Eq.~(\ref{ket_lambda1}), we get:
\begin{equation}
H'_{\lambda_1 0} = -E(t) \sum_{l_1} {U^{\lambda_1}_{l_1}}^* 
\mu^*_{l_1} 	\ ,
\end{equation}
where $U^{\lambda_1}_{l_1}$ is the unitary transformation from 
the non-interacting 
basis to the excitonic one, $l_1$ is the non-interacting two-particle label 
corresponding to the single-particle states $i_1$ and  $j_1$,
while $\mu_{l_1} \equiv \mu_{i_1j_1}$ is the single-particle 
dipole matrix element 
given in Eq.~(\ref{mu}).\\

For the biexcitonic absorption [see Eq.~(\ref{FGR3})]
the corresponding matrix elements are defined in 
Eq.~(\ref{H_lambda2_lambda1}).
Combining the explicit form of the carrier-light interaction Hamiltonian 
(\ref{H_prime}) with that of the excitonic state in 
Eq.~(\ref{ket_lambda1}) as well as of the biexcitonic state in 
Eq.~(\ref{ket_lambda2}), we obtain:
\begin{eqnarray}
H'_{\lambda_2 \lambda_1} &=& -E(t) \sum_{l_2} {U^{\lambda_2}_{l_2}}^* \cdot 
\nonumber \\
& &\cdot \left\{  
\left( U^{\lambda_1}_{i_1j_2} \mu^*_{i_2j_1}
+ U^{\lambda_1}_{i_2j_1} \mu^*_{i_1j_2}\right)
\left(2\delta_{i_1i_2,j_1j_2}-1\right)
- U^{\lambda_1}_{i_1j_1} \mu^*_{i_2j_2}
- U^{\lambda_1}_{i_2j_2} \mu^*_{i_1j_1}
\right\}
 \ ,
\end{eqnarray}
where again $U^{\lambda_N}_{l_N}$ is the unitary transformation from the 
non-interacting $N$-particle basis to the interacting one,
$\mu$ is the single-particle dipole matrix element, and $l_2 \equiv 
i_1j_1,i_2j_2$ is the generic label for the non-interacting two-pair basis.

\end{appendix}

\begin{figure} 
\caption{
Single-particle (dashed curve) and excitonic absorption spectra
(solid curve) in the field-free case ($F=0$). 
The inset shows how the exciton binding energy $\Delta{\cal E}$ is reduced 
as the in-plane electric field $F$ increases.
}
\label{f:sqd-abs1}
\end{figure} 
\begin{figure} 
\caption{
Single-particle (A) and excitonic absorption spectra (B) 
for an in-plane field $F=50$\,kV/cm. 
The excitonic spectrum in (B) ---apart from a rigid shift due to Coulomb 
interaction--- is now comparable to the single-particle one.
Here, numbers from 
$1$ to $4$ identify corresponding transitions in each spectrum.
}
\label{f:sqd-abs2}
\end{figure} 
\begin{figure} 
\caption{
Effective electron and hole charge distribution for the 
ground-state exciton in the field-free case. 
The three curves correspond to non-interacting (n.i.) as well as to 
Coulomb-correlated e-h pairs as indicated.
}
\label{f:sqd-cd1a}
\end{figure} 
\begin{figure} 
\caption{
Same as in Fig.~\protect\ref{f:sqd-cd1a}\protect\ but for an in-plane field
$F=50$\,kV/cm.\hfill\ 
} 
\label{f:sqd-cd1b}
\end{figure}
\begin{figure}
\caption{
Excitonic (solid curve) and biexcitonic optical response (dashed curve) 
for the antiparallel-spin configuration
in the presence of an in-plane electric field $F=30$\,kV/cm. 
The inset shows the biexcitonic splitting $\Delta{\cal E}$
as a function of the in-plane field $F$.
Notice that for $F=0$, the latter becomes very small
($\Delta{\cal E} = 0.7$\,meV), which is due to the special symmetry of 
the QD structure considered:
$\alpha_e = \alpha_h$ (see text).
} 
\label{f:sqd-abs3}
\end{figure}
\begin{figure}
\caption{
Excitonic (solid curve) and biexcitonic optical response (dashed curve) of 
a realistic QD structure
for the antiparallel-spin configuration
in the field-free case.
The inset shows the biexcitonic splitting $\Delta{\cal E}$
as a function of the in-plane field $F$.
Opposite to the symmetric case previously considered (see 
Fig.~\protect\ref{f:sqd-abs3}\protect),
in this more realistic case the spatial extension for electrons and 
holes as well as their Coulomb matrix elements 
are considerably different; this is the physical origin 
of the positive biexcitonic shift in the field-free case (see text).
}
\label{f:sqd-abs4}
\end{figure}
\begin{figure} 
\caption{
Effective electron and hole charge distributions 
for the ground-state exciton in the field-free case (solid curves)
as well as their 
difference (dashed curve). 
Due to the realistic QD parameters considered, 
the charge neutrality is violated: $\Delta f \neq 0$ (see text).
}
\label{f:sqd-cd2}
\end{figure} 
\begin{figure}
\caption{
Excitonic (solid curve) and biexcitonic optical response (dashed curve) of 
a realistic QD structure
for the parallel-spin configuration
in the field-free case.
The inset shows the biexcitonic splitting $\Delta{\cal E}$
as a function of the in-plane field $F$.
As can be seen from the spectra, the latter is now negative for 
$F=0$. However,
it becomes positive at high fields (see text).
}
\label{f:sqd-abs5}
\end{figure}
\begin{figure} 
\caption{
Schematic representation of the electron and hole charge distribution 
as well as of the confinement potential profile in our Coulomb-coupled 
QD structure.
The latter is tailored in such a way to allow for
an energy-selective creation/destruction of bound electron-hole pairs 
in dots $a$ and $b$.
Moreover, the inter-dot barrier width ($w \sim 50$\,\AA) is such to prevent
single-particle tunneling and at the same
time to allow for significant inter-dot Coulomb coupling (see text).
}
\label{f:cqd-str}
\end{figure}
\begin{figure} 
\caption{
Excitonic response of the array unit cell ($a+b$) in 
Fig.~\protect\ref{f:cqd-str}\protect\
for an in-plane field $F=75$\,kV/cm.
The Coulomb-correlated result in (B) is compared to the Coulomb-free one 
in (A).
}
\label{f:cqd-abs1}
\end{figure}
\begin{figure}
\caption{
Excitonic (solid curve) and biexcitonic spectrum (dashed curve)
for an in-plane field $F=75$\,kV/cm. 
Due to the well-defined polarization of our laser source, 
the structure in the biexcitonic spectrum (dashed curve) corresponds to 
the formation of an exciton in dot $b$ given an exciton in dot $a$. 
One obtains a similar structure in the biexcitonic spectrum, 
symmetrically blue shifted with respect to the excitonic transition in 
dot $a$, if one considers
as initial state an exciton in dot $b$. 
The biexcitonic shift $\Delta {\cal E}$
as a function of the in-plane field $F$ is also reported in the inset.
}
\label{f:cqd-abs2}
\end{figure}
\begin{figure}
\caption{
Effective potential $V_{eff}(x)$ (solid line)
as a function of the $x$ coordinate
for three different values of the external field $F$.
Here, the following parameters have been used:
$m_e=0.067 m_0$,
$m_h=0.34m_0$,
$\hbar\omega_e=30$\,meV,
$\hbar\omega_h=24$\,meV. 
The dashed line represents the parabolic part of $V_{eff}(x)$ and the 
dashed-dotted line the Coulomb term.
}
\label{f:Veff}
\end{figure}
\begin{figure}
\caption{
Biexcitonic shift $\Delta{\cal E}$ as a function of the in-plane field $F$. 
Here, the parameters used are the same as in 
Fig.~\protect\ref{f:Veff}\protect.
The squares represent exact numerical results, the solid line the
predictions of the model, the dashed line the
predictions of the model after setting $\Delta\omega=0$, and the dotted line
the results obtained by neglecting completely Coulomb interaction in the
wave function. 
The inset reports the behavior of $\Delta\omega/\omega_r$
and $\Delta x/d$ for the same 
range of applied fields.
}
\label{f:num_ana}
\end{figure}
\begin{figure}
\caption{
Plot of the parameter space available for designing the QD molecule used 
as quantum hardware in our QIC implementation.
This has been calculated using the proposed analytical model (see text).
}
\label{f:parsp}
\end{figure}
\begin{figure} 
\caption{
Time-dependent simulation of a two-qubit operation realizing the first 
prescription ($\vert 1,0 \rangle \to \vert 1,1 \rangle$)
for a CNOT logic quantum gate on dots $a$ and $b$ (see text).
Exciton populations $n_a$ and $n_b$ (upper panel) and diagonal 
density-matrix elements (lower panel)
as a function of time. The laser-pulse sequence is also sketched 
(central panel).
}
\label{f:tds1}
\end{figure}
\begin{figure} 
\caption{
Time-dependent simulation of a CNOT quantum gate transforming the factorized 
state $\vert 0,0 \rangle + \vert 1,0 \rangle$ 
into a maximally entangled state
$\vert 0,0 \rangle + \vert 1,1 \rangle$
for the coupled QD structure $a + b$ in Fig.~\protect\ref{f:cqd-str} 
(see text).
Exciton populations $n_a$ and $n_b$ (upper panel) and diagonal 
density-matrix elements (lower panel)
as a function of time. The laser-pulse sequence is also sketched 
(central panel).
}
\label{f:tds2}
\end{figure}


\begin{references}
\bibitem{T1-T2}
Decoherence processes are accounted for in our density-matrix formalism 
by means of a standard $T_1 T_2$ model: 
we employ a band-to-band recombination 
time $T_1 = 1$\,ns and we describe phonon-induced decoherence processes in 
terms of an exciton-phonon dephasing time $T_2 = 50$\,ps, 
which is compatible with the experimental values given in 
Ref.~\onlinecite{CCC}.
\bibitem{NN}
Due to the short-range nature of the exciton-exciton dipole interaction
[see Eq.~(\protect\ref{DelE}\protect)], 
we are allowed to consider 
nearest-neighbor coupling only.
Indeed, for the specific quantum hardware considered (see 
Fig.~\protect\ref{f:cqd-str}\protect),
in view of the ${1\over D^3}$ behavior and, more important, of their 
small oscillator strengths, second-neighbor couplings play no significant 
role.
\bibitem{screening}
The static dielectric constant $\varepsilon_0$ accounts for screening due 
to the crystal lattice only, i.e., due to the few-particle character of our 
system, free-carrier contributions should not be considered.
\end{references}
\end{document}